\documentclass[reprint,aps,prl,twocolumn,groupedaddress,amsmath,amssymb]{revtex4-1}
\usepackage{graphicx} 
\usepackage{dcolumn}  
\usepackage{bm}    
\usepackage{color}
\usepackage{dsfont}
\usepackage[lofdepth,lotdepth,caption=false]{subfig}
\DeclareMathOperator{\Tr}{Tr}
\usepackage{braket}

\begin{document}

\title{Efficient representation of long-range 
interactions in tensor network algorithms}
\author{Matthew J. O'Rourke}
\author{Zhendong Li}
\author{Garnet Kin-Lic Chan}
\affiliation{
   Division of Chemistry and Chemical Engineering, 
   California Institute of Technology, Pasadena, California 91125, USA
   }
   \begin{abstract}
We describe a practical and efficient approach to
represent physically realistic long-range interactions 
in two-dimensional tensor network algorithms via projected entangled-pair 
operators (PEPOs). We express the long-range interaction as a linear 
combination of  correlation functions of an auxiliary system 
with only nearest-neighbor interactions.
To obtain a smooth and radially isotropic 
interaction across all length scales, we map
the physical lattice to an auxiliary lattice of expanded size.
Our construction yields a long-range PEPO as a sum 
of ancillary PEPOs, each of
small, constant bond dimension. This representation enables
efficient numerical simulations with long-range
interactions using projected entangled pair states. 
\end{abstract}
%% is achieved
%% by interpolating multiple auxiliary system tensors
%% in between the original PEPO tensors.
%% This enables the 
%% For a given level of accuracy, it is argued that the 
%% number of terms in this sum is independent of the system 
%% size, which is the the key feature
%% that allows for the operator 
%% to be used effectively in numerical studies of
%% two-dimensional quantum systems that
%% employ both the finite and 
%% infinite projected entangled-pair ansatzes for the 
%% ground state wavefunction.

   \maketitle

\section{I. INTRODUCTION}
The accurate description of strongly correlated 
quantum many-body systems is a major 
challenge in contemporary physics. Nonetheless, 
some of the most intriguing macroscopic quantum phenomena,
such as high-temperature superconductivity and 
the fractional quantum Hall effect, 
arise from strong quantum correlations.
In recent years, tensor network states (TNS)
\cite{white1992dmrg,white1993dmrg,vidal2007mera,
changlani2009approximating,mezzacapo2009ground,
carleo2017solving}, including
matrix product states (MPS) 
\cite{ostlund1995thermodynamic,
fannes1992,fannes1994,schollwock2011density} 
and projected entangled-pair states (PEPS)
\cite{nishino1996corner,verstraete2004renormalization,
verstraete2006criticality,orus2014practical},
have emerged as promising classes
of variational states to numerically approximate
the low energy physics of correlated quantum systems 
 with area or near-area law physics. 
Their power stems from systematically improvable accuracy 
through increasing the tensor bond dimension $D$
\cite{corboz2016improved}, and 
the $O(A)$ linear complexity of the associated 
algorithms with respect to the system size $A$ (under assumption of
contractibility of the underlying tensor network, as is common in many physical applications, using approximate contraction methods
~\cite{verstraete2004renormalization,murg2007variational,
nishino1996corner,orus2009simulation,PhysRevLett.113.046402,
levin2007tensor,evenbly2015tensor}.)

One promising application of TNS is to 
accurate calculations of electronic structure of realistic materials. 
While the electronic structure Hamiltonian 
can be represented in multiple ways
\cite{martin2004electronic,szabo1996modern,
white1999ab,chan2016matrix}, 
the simplest -- and the one of interest in this work -- is a
real-space grid formulation
\cite{stoudenmire2012one,wagner2012reference,
stoudenmire2017sliced,dolfi2012multigrid,narbe2018},
\begin{gather}
\hat{H} = -t\sum_{<i,j>}(a_{i\sigma}^\dagger 
a_{j\sigma}+h.c.)+\sum_{i}v^{ne}_i n_i
+\hat{V}^{ee},\nonumber \\
\hat{V}^{ee} =
\sum_{i}v_{ii}^{ee}n_{i\alpha}n_{i\beta}+
\sum_{i<j}v_{ij}^{ee}n_{i}n_{j},
\label{eqn:ESham}
\end{gather}
where $i,j$ label lattice sites, 
$\sigma\in \{ \alpha,\beta\}$ labels spin, 
$t$
is the kinetic energy matrix element,
%central difference coefficient of the 
%hopping term,
and $a^{\dag}$, $a$, and $n$ are
fermion creation, annihilation, and number 
operators, respectively. As the spacing between 
grid points ($h$) goes to zero, the parameters
scale as $t \propto h^{-2}$ 
and $v_{ij}^{ee} \propto h^{-1}$; 
these become exact representations
of $-\frac{1}{2}\nabla^2$ and the
continuum Coulomb potential $1/r_{ij}$ with
$r_{ij} \triangleq |\textbf{r}_i - \textbf{r}_j|$
\cite{wagner2012reference,dolfi2012multigrid}.
This simple
form of the electronic structure Hamiltonian is 
especially suited to
TNS algorithms as the Coulomb interaction is a pairwise
operator as opposed to a general quartic operator 
when using a non-local basis, 
and Eq.~(\ref{eqn:ESham}) can be viewed as an extended Hubbard model with long-range terms.
Ground states of such grid Hamiltonians have been computed in 1D 
using MPS and the density matrix renormalization
group (DMRG), 
yielding near exact electronic structure benchmarks 
for small lattice spacings~\cite{white1992dmrg,
white1993dmrg,stoudenmire2012one}.
In principle, this success in 1D should be extensible
to 2D and 3D by using PEPS instead of MPS, and 
would then provide a route to simulate arbitrarily
complex electronic structure problems with 
arbitrarily improvable accuracy.

However, current state-of-the-art PEPS 
applications to physical problems have not yet advanced beyond local
lattice models in 2D~\cite{PhysRevX.4.011025,PhysRevLett.113.046402,
PhysRevLett.112.147203,PhysRevB.91.064415,
PhysRevB.93.060407,Zheng1155}. There are two principal complications. The first
is that long-range interactions can in principle lead to increased entanglement, and even volume-law entanglement, that would be difficult or impossible to capture with a PEPS with a finite bond dimension. Fortunately, in applications of the density matrix renormalization group using the Coulomb interaction (for example, to electronic structure) it is seen that the increase in entanglement is modest and volume law entanglement is not observed~\cite{fano1998density,white1999ab,chan2002highly,chan2011density,stoudenmire2012one,stoudenmire2017sliced}. The second complication 
is simply the increased cost of all operations when long-range
interactions are considered, even for a fixed bond dimension.
To see the basic
challenge, consider the evaluation of the energy expectation
value: for a Hamiltonian with localized interactions,
the number of terms in a standard term-by-term calculation
scales linearly with 
the size of the system, $O(A)$. However,
for a Hamiltonian with long-range interactions, 
the number of terms scales like $O(A^2)$, which is
prohibitively expensive in two (or higher) dimensions, 
as we take the continuum limit. 
Alternatively, one might try to use an exact tensor network 
operator, or projected entangled pair operator (PEPO), 
to represent the long-range interaction~\cite{pirvu2010matrix}, 
avoiding the explicit term-by-term evaluation. However, the 
exact PEPO representation for arbitrary long-range interactions 
in 2D has a bond dimension that scales as $O(A^{1/4})$,
causing the overall cost to compute expectation values to 
scale as $O(A^2)$~\cite{frowis2010tensor}.

In 1D, the increased computational cost of long-range interactions 
can be eliminated if they are smooth and decaying. In this
case one can approximate the exact matrix product operator (MPO)
by a compressed MPO of constant bond dimension 
$D$ that generates a sum of exponential interactions,
and smoothly decaying interactions can be approximated well by such sums~\cite{pirvu2010matrix,crosswhite2008finite,
crosswhite2008applying}.
Exponential interactions in MPOs arise naturally from
the matrix product structure, which also gives rise 
to the exponential decay of two-point
correlation functions in MPS. Extending the correlation 
function analogy to 2D leads
to an efficient representation of long range interactions 
in 2D when their form exactly
coincides with the correlation function of a 2D lattice model. 
This was demonstrated in Ref.~\cite{pirvu2010matrix}, 
which constructed a compact pair interaction PEPO
whose interaction potential was given by the
critical 2D Ising correlation function.

\begin{figure}[t]
\begin{center}
\subfloat[]{
\includegraphics[width=0.42\textwidth]{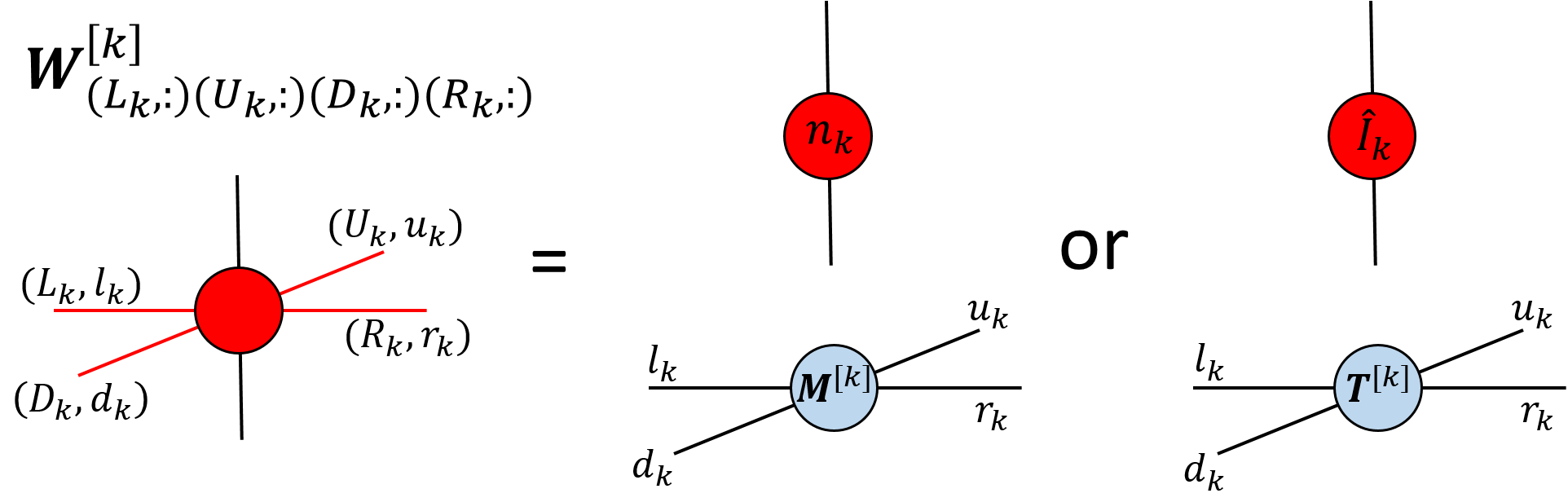}
\label{fig:cfpepo}}\\
\subfloat[]{
\includegraphics[width=0.2\textwidth]{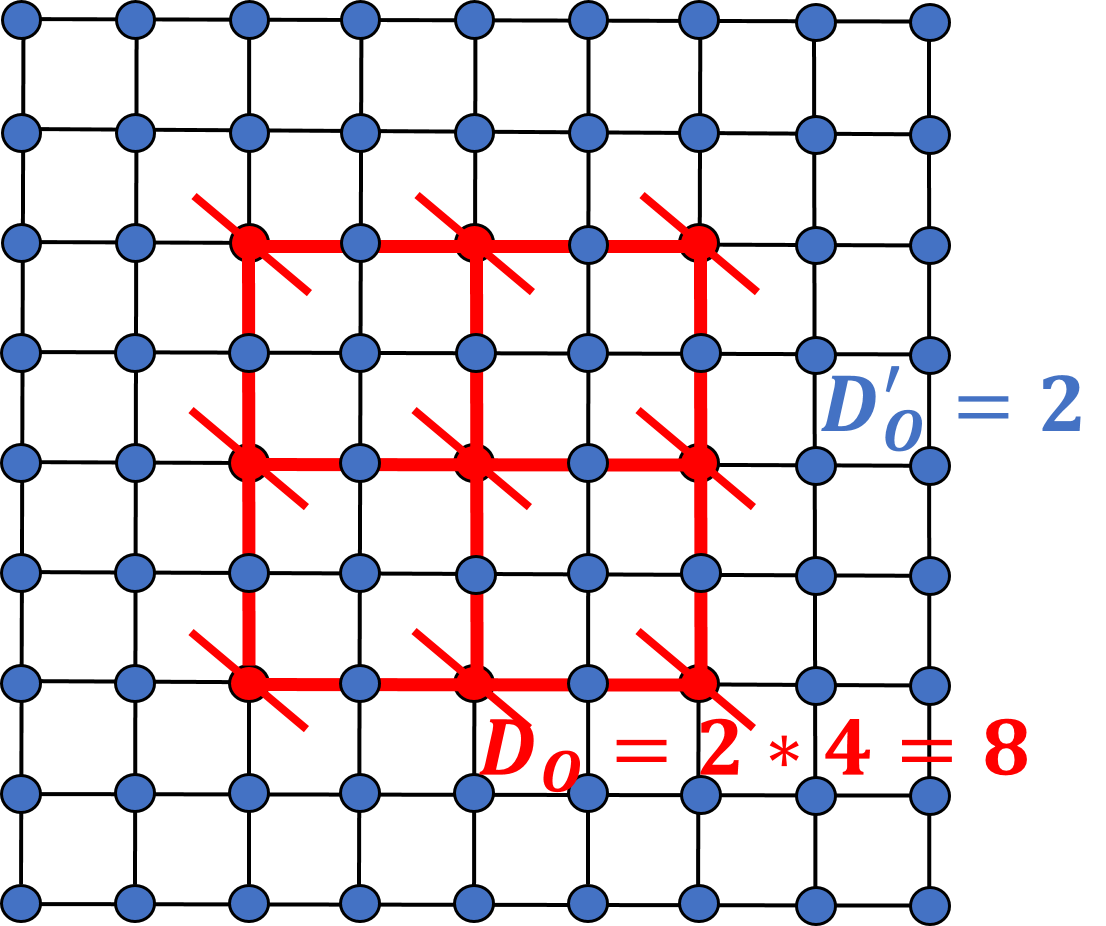}
\label{fig:fullPEPO}}
\subfloat[]{
\includegraphics[width=0.2\textwidth]{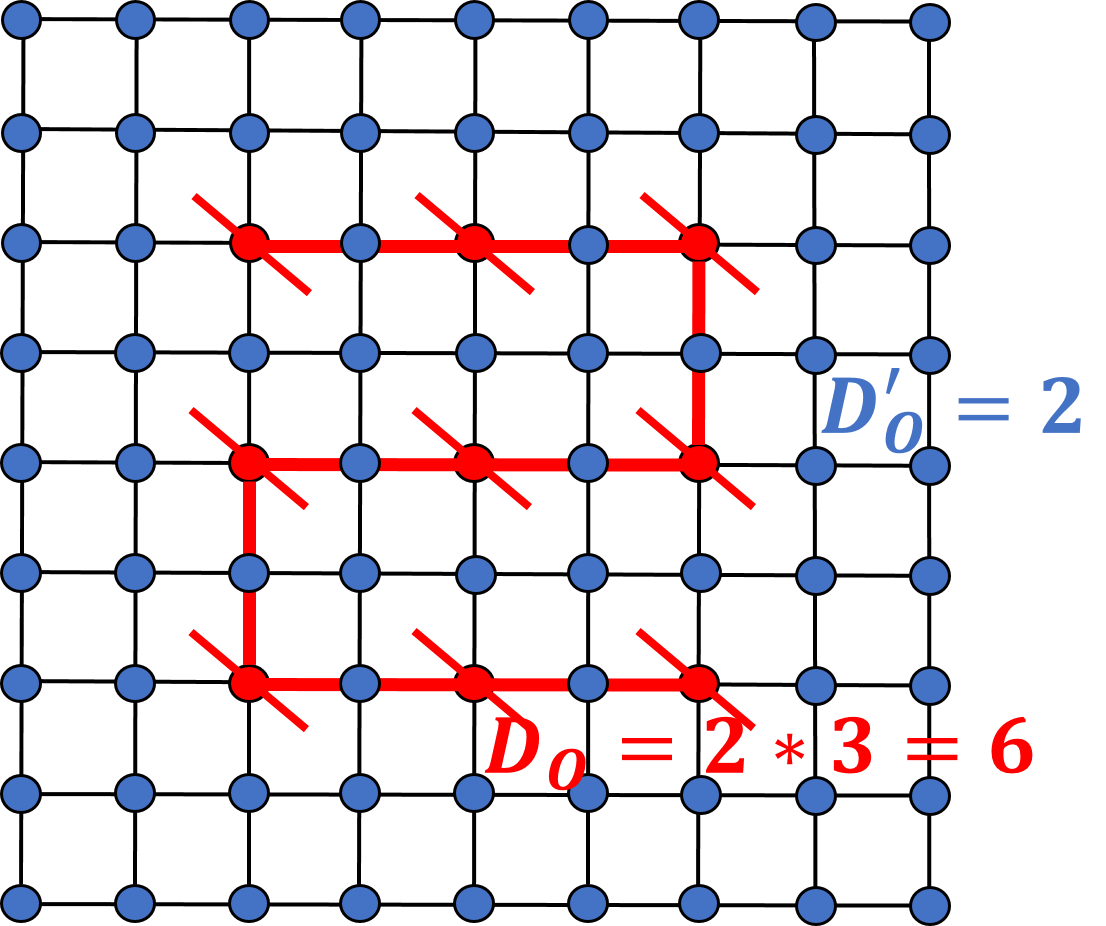}
\label{fig:snakePEPO}}\\[-0.8ex]
\caption{(a) The
construction of the nonzero parts of the CF-PEPO 
tensor $\mathbf{W}^{[k]}$ via the coupling of
the finite state machine (FSM) tensor (red) with the 
Ising correlation function tensors (blue). 
Note that here the physical indices of $\mathbf{W}^{[k]}$
are explicitly shown, whereas they are suppressed
in Eq. \eqref{eqn:PEPO}.
(b)-(c) Two possible constructions of the
long-range PEPO for a 3x3 physical 
system with 1 fictitious Ising site (blue) in 
between adjacent physical sites (red) and a 2 
site buffer to help mitigate boundary effects 
in the encoding of the potential. Black bonds 
are $D'_O=2$ and red bonds are $D_O=8$ (b) and 6 (c).
}
\label{fig:CF-PEPO}
\end{center}
\end{figure}

Building on these ideas, in this work we describe how 
general long-range interactions in two dimensions, 
including the Coulomb interaction,
can be efficiently encoded as a sum of low rank correlation 
function valued PEPOs.
Although superficially similar to the problem of approximating
a smooth interaction in 1D by a sum of exponentials, 
additional complications arise in two dimensions
because physical interactions possess different analytic 
properties from two-point correlation functions 
on the same lattice.
For example, the Coulomb interaction is radially isotropic 
at all distances, while 
the two-point lattice correlation functions are isotropic 
only at large distances due to the lattice discretization.
We show how to overcome these and other difficulties
by introducing an expanded auxiliary lattice, and demonstrate 
the effectiveness of the representation
in a ground-state finite PEPS simulation of a 2D spin model 
with Coulombic Heisenberg interactions. Although we specifically 
treat only the Coulomb interaction and two dimensions in our 
numerical examples, our arguments naturally extend to 
representing smooth and radially isotropic 
interactions in any dimension.

\section{II. CORRELATION FUNCTION VALUED PEPOS}
We first define correlation function valued PEPOs (CF-PEPOs),
which are central to this work. As motivation, we
recall the construction of MPOs for smooth interactions 
approximated by sums of exponentials. This is usually done in
the language of finite state machines (FSM), where the 
MPO is viewed as an operator valued MPS, and the
incoming and outgoing bonds of each MPO tensor are
interpreted as machine states~\cite{crosswhite2008finite,
crosswhite2008applying}.
An FSM can encode an exponentially decaying interaction 
strength $e^{-\lambda r_{ij}}$ via
a single non-zero element in each MPO tensor with value 
$e^{-\lambda}$, that gets multiplied along the lattice 
as long
as the FSM stays in a specified state. The pairwise 
operator $\sum_{i<j} e^{-\lambda r_{ij}} n_i n_j$ can then
be represented by an MPO with bond dimension 3, with the 
two additional states in the FSM acting to combine the 
exponential scalar values with the operators $n_i n_j$.
The construction can be extended to the general 1D interaction
$\sum_{i<j} V(r_{ij}) n_i n_j \approx 
\sum_{i<j} \sum_{t=1}^{N_t} 
c_t e^{-\lambda_t r_{ij}} n_i n_j$
by introducing additional states for each of the $N_t$ 
exponential decays, for a total MPO bond dimension of $N_t+2$
(or alternatively, $N_t$ MPOs of bond dimension 3).
However, while this representation is natural in 1D, 
its direct extension to 2D is not.
This is because multiplying the element $e^{-\lambda}$ 
along any single FSM path between two sites $i$ and $j$
creates an exponentially decaying strength as a 
function of the Manhattan distance $|x|+|y|$, not
the desired Euclidean distance $(x^2+y^2)^{1/2}$, 
as the elements are multiplied out along 
the grid lines \cite{frowis2010tensor}.

A different starting point, that is more natural in 
higher dimensions, is to consider scalar interaction strengths
generated by the two-point correlation function
$\langle  o(\textbf{r}_i) o(\textbf{r}_j) \rangle_\beta$ 
of a classical model at inverse temperature $\beta$. 
We term the PEPO for
the operator $\sum_{i<j} \langle  o(\textbf{r}_i) 
o(\textbf{r}_j) \rangle_\beta n_i n_j$, a
correlation function valued PEPO (CF-PEPO). 
Using a classical model
with local interactions yields a CF-PEPO with
low bond dimension, as noted in  Ref.~\cite{pirvu2010matrix}.
As a concrete example, consider the spin-spin correlation 
function $\langle \sigma_i \sigma_j \rangle$ of the
2D Ising model, which has the Hamiltonian
$H=-\sum_{\langle m,n\rangle }\sigma_m\sigma_n$, 
$\sigma \in\{+1,-1\}$. For two given points on
the lattice $i$ and $j$, this correlation
can be exactly represented by the Ising
PEPS with $D=2$ \cite{verstraete2006criticality,
zhao2010renormalization}, viz.,
\begin{equation}
\langle \sigma_i \sigma_j \rangle_\beta  = \frac{1}{Z}
\Tr\left(\prod_{k\ne i,j} T^{[k]}_{l_k u_k d_k r_k} 
M^{[i]}_{l_iu_id_ir_i}M^{[j]}_{l_ju_jd_jr_j}
\right).
\label{eqn:ising_corr}
\end{equation}
Here $Z=\Tr\prod_{k} \mathbf{T}^{[k]}$ is the partition
function and the tensors $\mathbf{T}$ and $\mathbf{M}$ are
the local tensors of the PEPS off and on the correlation function
sites, respectively. These tensors are 
obtained from the eigenvalue decomposition 
$X = U \lambda U^{\mathrm{T}}$
of the familiar 2$\times$2 Ising model transfer 
matrix $X_{ij} = \exp((-1)^{\delta_{ij}+1}J\beta)$,
which encodes the local terms of the partition function
for a pair of nearest neighbor spins \cite{baxter1982exactly}.
In tensor network language, these $X$ matrices
would be placed on each bond of the square lattice.
In order to create a local tensor network description
of the system, we define the ``square root" of this transfer operator
as $P = U \sqrt{\lambda} U^{\mathrm{T}}$,
and define the local tensors as
$T_{ludr} = \sum_a P_{la} P_{ua} P_{ad} P_{ar}$ and
$M_{ludr} = \sum_{ab} P_{la} P_{ua} P_{da} \sigma^z_{ab} P_{br}$,
where $\sigma^z$ is the standard Pauli matrix. 

To obtain the Ising CF-PEPO, we combine the tensors $\mathbf{T}^{[k]}$, $\mathbf{M}^{[k]}$ of the Ising PEPS at each site with (translationally invariant) tensors $\mathbf{Y}^{[k]}$ of a PEPO for the interaction $\sum_{i<j} n_i n_j$. As demonstrated in a general fashion in~\cite{frowis2010tensor} based on work in \cite{crosswhite2008finite}, the $\mathbf{Y}^{[k]}$  tensors can be obtained by a FSM construction in 2D, where each element of the tensor $Y_{L,U,D,R}$ at a given site corresponds to a specific local state of the FSM and returns a specific local operator $\{0, \hat{I}, n\}$.
%the main difference from similar 1D constructions being the
%larger number of possible paths that the FSM is able 
%to traverse, leading to a larger number of rules. For this reason, the bond dimension of $\mathbf{Y}^{[k]}$ is 4, compared to a bond dimension of 3 for the analogous MPO.
%obtain the Ising CF-PEPO
%by combining its tensors with the tensors of a FSM that
%generates all pairwise (and only pairwise) interaction terms 
%$n_i n_j$ on the square lattice. 
%A very general form
%of this type of operator valued FSM was
%derived in \cite{frowis2010tensor}, based on work in
%\cite{crosswhite2008finite}. The essential ideas are similar to the use of FSMs in 1D to generate MPOs, however in 2D the . The FSM In the
%end, this set of rules defines an operator valued PEPS (thus,
%a PEPO) composed of the same tensor $Y_{L,U,D,R}$ at each site,
%where the specific values of the %indices $(L,U,D,R)$ corresponds
%to a specific local state of the FSM, and thus returns a 
%specific local operator $\{0, \hat{I}, n\}$. 
The Ising CF-PEPO tensors are then formed by a selective direct product between $\mathbf{Y}^{[k]}$, $\mathbf{T}^{[k]}$, 
and $\mathbf{M}^{[k]}$,
\begin{gather}
\sum_{i<j}\langle \sigma_i\sigma_j\rangle_{\beta} 
n_{i}n_{j} = \Tr\left(\prod_k 
W^{[k]}_{(L_k,l_k)(U_k,u_k)(D_k,d_k)(R_k,r_k)}\right),\nonumber\\
\mathbf{W}^{[k]}_{(L_k,:)(U_k,:)(D_k,:)(R_k,:)} = 
Y^{[k]}_{L_k, U_k, D_k, R_k} \otimes \mathbf{T}^{[k]} 
~\mathrm{if} ~Y = \hat{I}_k, \nonumber\\
\mathbf{W}^{[k]}_{(L_k,:)(U_k,:)(D_k,:)(R_k,:)} = 
Y^{[k]}_{L_k, U_k, D_k, R_k} \otimes \mathbf{M}^{[k]} 
~\mathrm{if} ~Y = n_k, \nonumber\\
\mathbf{W}^{[k]}_{(L_k,:)(U_k,:)(D_k,:)(R_k,:)} = 
0
~\mathrm{if} ~Y = 0. 
%\mathbf{W}^{[k]}_{(L_k,:)(U_k,:)(D_k,:)(R_k,:)}
%\triangleq
%\mathbf{X}^{[k]}, \nonumber\\ \mathbf{X}^{[k]} \in 
%\{ 0, \mathbf{T}^{[k]} \otimes \hat{I}_k, \mathbf{M}^{[k]} \otimes n_k\}.
%%P^{[k]}_{L_kU_kD_kR_k}X^{[k]}_{l_ku_kd_kr_k}O^{[k]}_{n_kn'_k},
\label{eqn:PEPO}
\end{gather}
Here $\mathbf{W}^{[k]}$ (Fig. \ref{fig:CF-PEPO}(a))
is the operator valued tensor 
in the Ising CF-PEPO and $(L_k,l_k)$ is a composite index 
of the bond $L_k$ for the 2D FSM and
the bond $l_k$ of the Ising PEPS. Note that the selective
direct product can be formed unambiguously
due to the $1:1$ correspondence between possible
states of $\textbf{Y}^{[k]}$ and the Ising PEPS tensors
$\mathbf{M}^{[k]}$ and $\mathbf{T}^{[k]}$.

Since the FSM tensors $\mathbf{Y}^{[k]}$ only need to 
encode the two operators $n_i n_j$
and contain no information about the distance between
them, there is some flexibility in the possible
topologies of the FSM (see Fig.
\ref{fig:CF-PEPO}). The snake geometry in (c) has a
significantly reduced computational complexity compared
to the original FSM from \cite{frowis2010tensor} shown in (b),
and it also imposes an ordering that allows for a 
simple way to include fermionic statistics (via
Jordan-Wigner strings) at the operator level,
eliminating the need for swap gates in
fermionic PEPS
\cite{corboz2010simulation}. The full specifications
for constructing the tensors $\mathbf{Y}^{[k]}$ 
according to both FSM geometries are given in Appendix
A. As an important note, both of these
constructions are compatible with existing iPEPS 
\cite{jordan2008classical} algorithms.

\section{III. CF-PEPOS AND THE AUXILIARY LATTICE} 

\begin{figure}[t]
\begin{center}
\subfloat{
\includegraphics[width=0.47\textwidth]{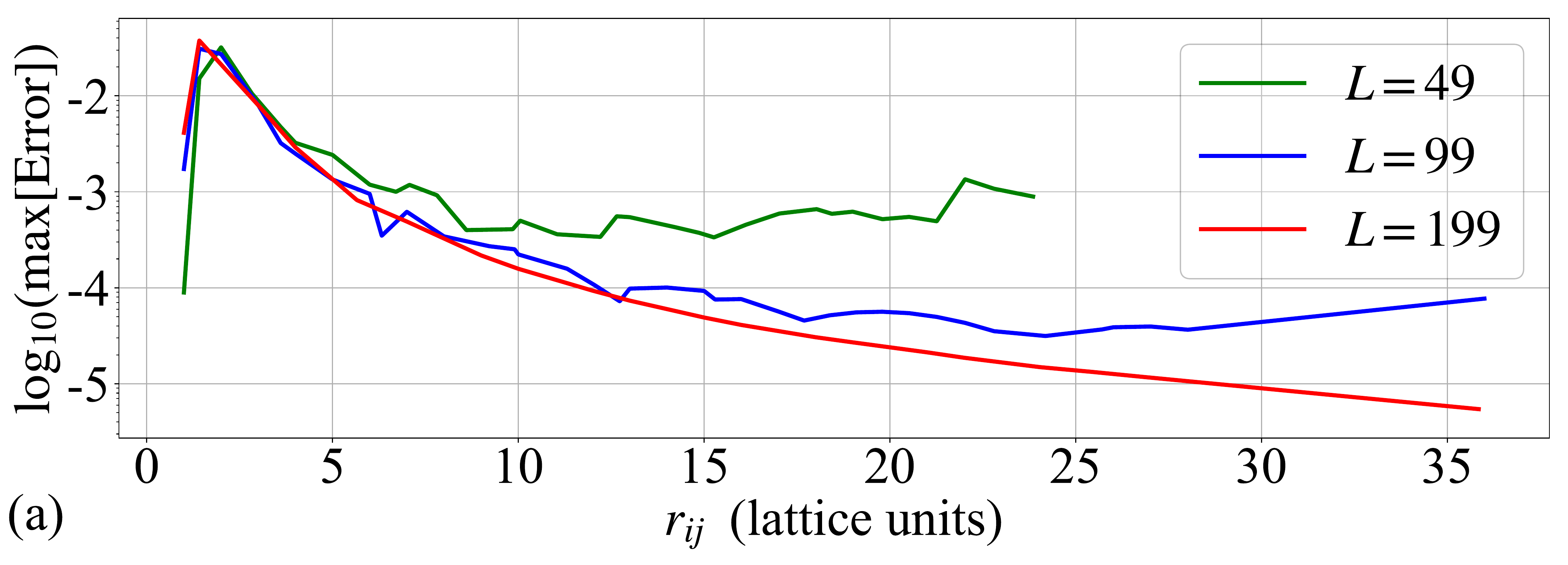}
\label{fig:isotropy1}}\\[-0.8ex]
\subfloat{
\includegraphics[width=0.47\textwidth]{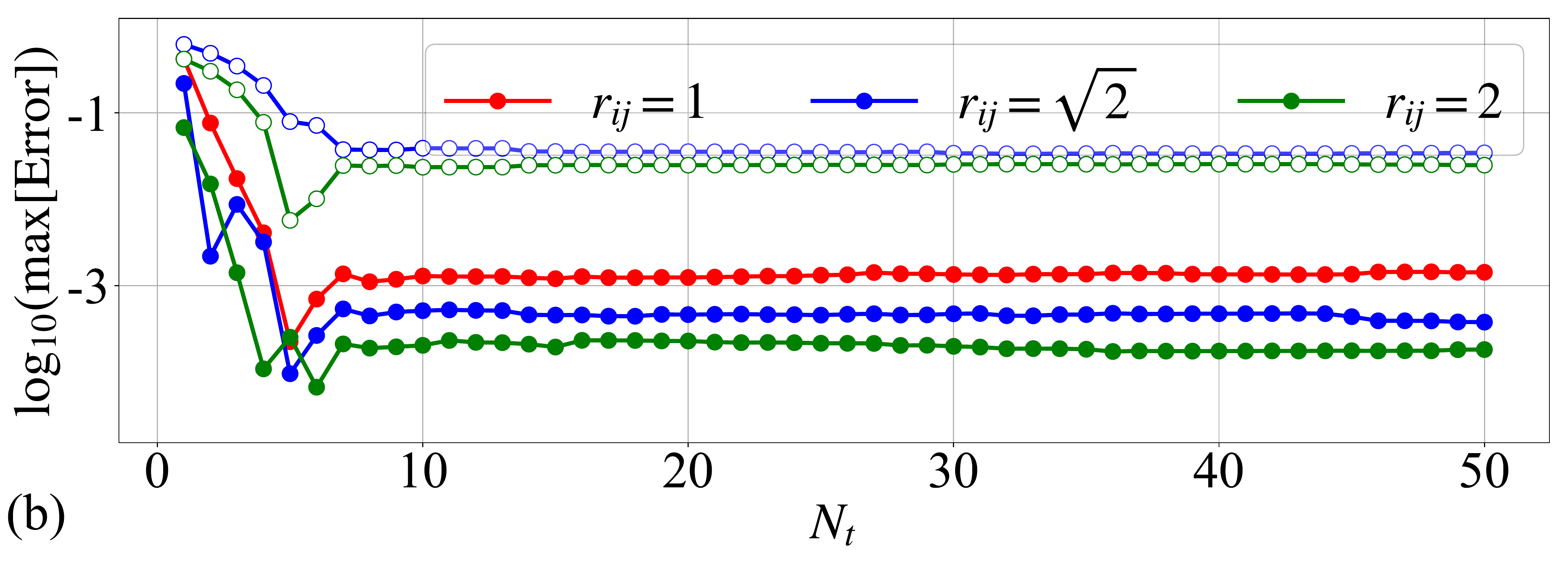}
%\caption{}
\label{fig:isotropy2}}\\[-0.8ex]
\subfloat{
\includegraphics[width=0.47\textwidth]{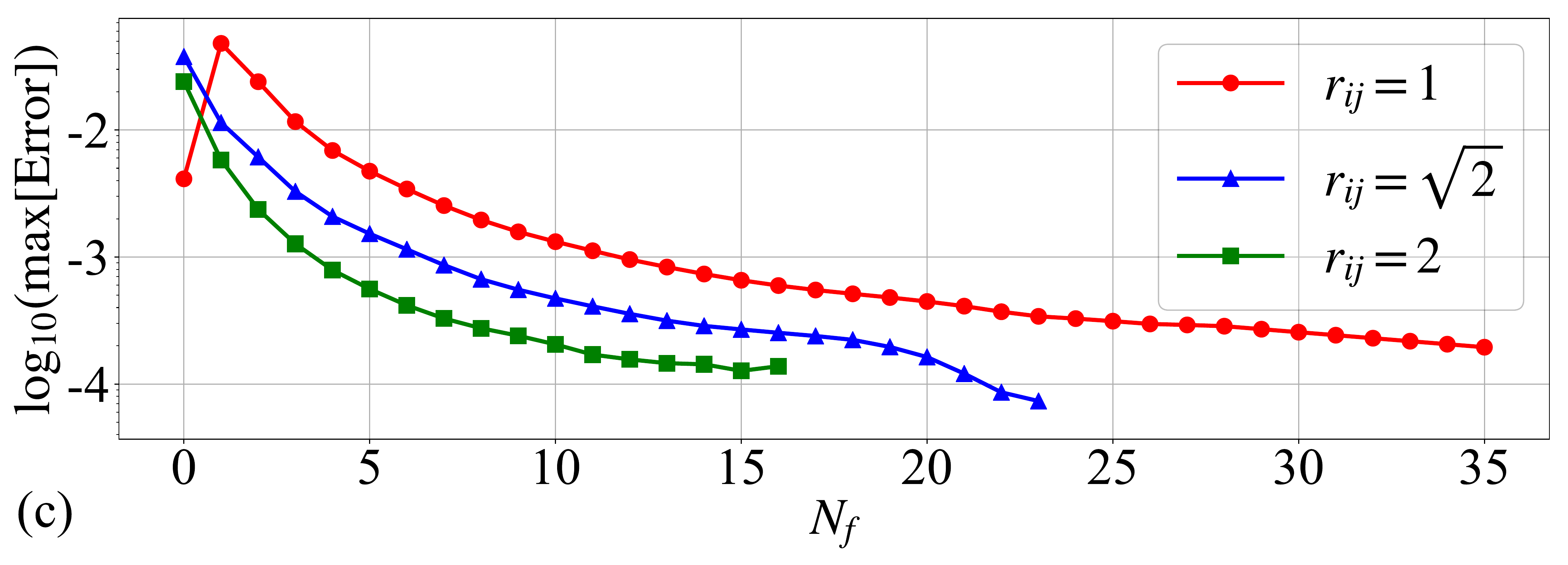}
%\caption{}
\label{fig:isotropy3}}
\\[-2ex]
\caption{Convergence properties of 
Coulomb fitting. For all plots $r_{ij}=0$ 
is the central point on the lattice. (a) The 
upper envelope of 
$\left| V_{\mathrm{fit}}(r_{ij}) - 1/r_{ij} \right|$
obtained with $N_t=12$, $r_{ij}=R_{ij}$,
a least squares weight
function of $r_{ij}^{1.5}$, and 
Ising model lattices with 
different side lengths $L$. 
The fits were performed 
on a disc with radius equal to the maximum
$r_{ij}$ displayed for a given curve. 
(b) and (c): The maximum fitting error
$\left| \tilde{V}_{\mathrm{fit}}^{[N_f]} - 1/r_{ij} \right|$
at selected values of $r_{ij}$ as functions of
$N_t$ (b) and $N_f$ (c). In (b), the open circles
correspond to $N_f=0$ and the closed circles to
$N_f=10$. In (c), $N_t=12$.
The fits in (b) and (c) were performed on discs 
of radius $r_{ij}=36$
with $L=199$ and a weight function of $r_{ij}^{1.5}$.
}
\label{fig:isotropy}
\end{center}
\end{figure}

Using the above arguments, we might now consider approximating the form of 
a physical, smooth, and isotropic interaction $V(r_{ij})$ by a sum
of $N_t$ lattice correlation functions at different temperatures, 
$V(r_{ij}) \approx V_{\mathrm{fit}}(r_{ij})=
\sum_{t=1}^{N_t} c_t f_{\beta_t}(r_{ij})$ 
[$f_{\beta_t}(r_{ij}) \triangleq \langle o(\mathbf{r}_i) 
o(\mathbf{r}_j)\rangle_{\beta_t}$],
giving the interaction operator as a sum of CF-PEPOs. 
In Fig. \ref{fig:isotropy}(a)
we show the maximal absolute error in a 
direct fit %(i.e. $R_{ij}=r_{ij}$)} 
of $1/r_{ij}$ using Ising correlation
functions on an $L$x$L$ lattice.
For large $r_{ij}$, the maximal error (at a given radius)
can be seen to converge rapidly, with a fitted convergence 
rate of $\sim O(r_{ij}^{-2.7})$
(Fig. \ref{fig:isotropy}(a)), showing we can easily capture
the long distance behavior of the Coulomb potential 
that is sampled at large system sizes.
However, for small $r_{ij}$, the maximal 
errors are much larger, and the expansion does {\it not} converge 
even with very many terms, as seen in Fig. \ref{fig:isotropy}(b). 
This is because the lattice discretization of 
the correlation functions
prevents radial isotropy in the basis 
$\{ f_{\beta_t} \}$ at short lattice distances. 
In addition, for finite lattices, 
boundary effects also cause errors in the 
isotropy and translational invariance.
%For small values of $r_{ij}$, the radial anisotropies 
%in the basis  
%cause the fit to always be poor, even with increasing 
%numbers of fitting functions, as seen in 
%{\color{red} 
%Here $R_{ij} = r_{ij}$, but we are making 
%a suggestive distinction
%between the radial coordinate of the physical potential,
%$R_{ij}$, and the natural radial coordinate in the 
%underlying correlation function representation, $r_{ij}$,
%which is fixed by the Ising model lattice.}
%We make this distinction because, taking
%$V(R_{ij})=1/R_{ij}$ and $f_{\beta_t}(r_{ij})$ as the 
%Ising correlation function as examples, we see that
%a direct expansion is not efficient because, 

The short distance anisotropy error can be remedied
by representing the isotropic physical interaction by
correlation functions generated on an {\it expanded auxiliary lattice} with additional
``fictitious" sites.
The physical distance $r_{ij}$ (on the original lattice) maps to the 
expanded distance $R_{ij} = (N_f+1) r_{ij}$ on the 
auxiliary lattice ($N_f$ denotes the number of
fictitious sites added 
to the sides of one unit square on the original
lattice). This gives us a rescaled potential that is easier to fit at small $r_{ij}$,
\begin{equation}
\tilde{V}^{[N_f]}_{\mathrm{fit}}(r_{ij}) \triangleq (N_f+1) 
V_{\mathrm{fit}} (R_{ij}) = (N_f+1) \sum_{t=1}^{N_t} c_t f_{\beta_t}(R_{ij}),
\label{eq:scaleFit}
\end{equation}
where the specific rescaling in Eq.~\eqref{eq:scaleFit} has been shown for 
the Coulomb potential.
Choosing a sufficiently large expansion factor $N_f$ ensures
that the fitting basis becomes isotropic up to an error $\epsilon$, and
the radial fit can then be performed to increasing
accuracy with increasing $N_t$ up to a similar $\epsilon$.
Further, choosing a suitably 
large side length of the auxiliary lattice buffering the physical region
also removes the boundary effects in a finite lattice simulation.

%thus giving $V(r_{ij}) \approx V_{\mathrm{fit}}(r_{ij})=
%\sum_{t=1}^{N_t} c_t f_{\beta_t}(R_{ij})$.
%Choosing a sufficiently large expansion factor ensures
%that the basis is isotropic up to an error $\epsilon$, and
%the radial fit can then be performed to increasing
%accuracy with increasing $N_t$ up to a similar $\epsilon$.
%Further, choosing a suitably 
%large side length of the auxiliary lattice buffering the physical region,
%also removes the boundary effects in a finite lattice simulation.

In Figs.~\ref{fig:isotropy}(b)-(c)
we show the behavior of the maximal error in fitting
$\tilde{V}_{\mathrm{fit}}^{[N_f]}(r_{ij})$
to $1/r_{ij}$ for several values of $r_{ij}$,
as a function of both the number 
of fictitious sites $N_f$ and fitting terms $N_t$.
They demonstrate that for 
$N_f=10$ and a modest $N_t=8$, we are able 
to obtain a maximum error of $10^{-3}$ with
$\tilde{V}_{\mathrm{fit}}^{[10]}(r_{ij})$.
In Fig. \ref{fig:isotropy}(c), note that the $r_{ij}=1$
curve (i.e. the maximal error curve) converges
as $\sim (N_f+1) O(N_f^{-2.7}) \propto N_f^{-1.7}$ due to the
rescaling factor in Eq. \eqref{eq:scaleFit}. 
Thus by further increasing $N_f$ 
the error can be continually
decreased.

Up to this point in this section, we have implicitly
considered $r_{ij}$ only on the unit lattice, i.e., 
$r_{ij} \triangleq |(x,y)_i - (x,y)_j|; x,y \in \mathbb{Z}$,
which is to say that the lattice spacing $h=1$. 
In addition to the above discussion of increasing $N_f$ to
reduce the fitting error for a fixed spacing $h=1$, an alternative
(but equivalent) viewpoint is that $N_f$ can be increased
to maintain a given maximal error in 
the potential as $h \to 0$.
Precisely, the maximal error in the new potential will
occur at the new shortest physical distance,
$V(h) = h^{-1} \tilde{V}^{[N_f]}_{\mathrm{fit}}(1)$.
The error at this point $\epsilon(V(h))$ scales as 
$\epsilon \propto h^{-1} N_f^{-1.7}$, which reveals
that $N_f$ must increase as $N_f \propto h^{-1/1.7} = h^{-0.59}$
in order to maintain the level of error originally incurred 
at the point $\tilde{V}^{[N_f]}_{\mathrm{fit}}(1)$ (for $h=1$).

In summary, the full CF-PEPO is obtained by coupling
the FSM of the operators (either in the snake form, or the full 2D FSM) to the Ising CF-PEPS on an expanded lattice
as specified by Eq.~\eqref{eq:scaleFit}, and as shown in Fig. 
\ref{fig:CF-PEPO}(b)-(c). The total error of the fit is controlled by the expansion
parameter $N_f$ and the number of terms $N_t$. For the Coulomb interaction and
a desired accuracy, $N_t$ is only weakly dependent on the 
physical lattice discretization and system size. 
%% For system sizes $r_{ij} \in [1,\tilde{R}]$ that are
%% not unreasonably small,
%% the behavior of these small-$r_{ij}$ errors is 
%% unaffected by the extent of the outer boundary $\tilde{R}$.
%% Thus, the total error of the fit is strongly
%% dominated by terms for which the saturated value
%% of $N_t$ is small and fixed (independent of $\tilde{R}$).
%% It can therefore be concluded that, for a given
%% desired level of accuracy in fitting $1/R_{ij}$,
%% the number of terms $N_t$ in
%% $\tilde{V}_{\mathrm{fit}}^{[N_f]}(R_{ij})$
%% is independent of, or at most weakly dependent on, 
%% system size.
This is similar to what is observed in MPO fits 
in one dimension \cite{crosswhite2008applying,
crosswhite2008finite,pirvu2010matrix,stoudenmire2012one}
as well as analytical work on exponential fits 
of the Coulomb operator in 2D~\cite{constNumFitFuncs}.

%% For system sizes $r_{ij} \in [1,\tilde{R}]$ that are
%% not unreasonably small,
%% the behavior of these small-$r_{ij}$ errors is 
%% unaffected by the extent of the outer boundary $\tilde{R}$.
%% Thus, the total error of the fit is strongly
%% dominated by terms for which the saturated value
%% of $N_t$ is small and fixed (independent of $\tilde{R}$).
%% It can therefore be concluded that, for a given
%% desired level of accuracy in fitting $1/R_{ij}$,
%% the number of terms $N_t$ in
%% $\tilde{V}_{\mathrm{fit}}^{[N_f]}(R_{ij})$
%% is independent of, or at most weakly dependent on, 
%% system size.

%As can be seen in Fig.~\ref{fig:isotropy}(a),
%for {\color{red} $r_{ij}>2$}
%the total error in the fit will
%be strongly dominated by the errors
%at the few smallest values of $R_{ij}$. 

\section{IV. COMPUTATIONAL COST}

We now consider the evaluation of a finite PEPS expectation value
for a PEPS of bond dimension $D_S$ and 
an Ising CF-PEPO of bond dimension $D_O$. 
To define the computational cost, we must choose
an approximate contraction scheme. Here we use
a simple generalization of the ``optimized'' contraction scheme
proposed in Ref. \cite{xie2017optimized} to include a PEPO.
Using the full 2D FSM
(Fig. \ref{fig:CF-PEPO}(b)), the CF-PEPO has bond
dimension $D_O=8$ for the bonds emanating from the 
physical sites and $D'_O=2$ for bonds that only connect
fictitious sites, and the leading contraction cost can be derived to be
$N_t [O(A \chi^3 D_O^3) + O(A N_f \chi^3 D_O^{'2} D_O)+ 
O(A N_f^2 \chi^3 D_O^{'3}) + O(A \chi^3 D_S^3) +
O(A N_f \chi^3 D_O^{'2} D_S)]$,
where $\chi$ is the maximum bond dimension appearing
in the approximate contraction scheme and can be taken as $\chi \sim D_S^2D_O$. 
For the snake FSM construction (Fig. \ref{fig:CF-PEPO}(c))
$D_O=6$ instead of 8, and the physical PEPO tensors only
have two large bond dimensions instead of four. This reduces the overall scaling to
$N_t [O(A \chi^3 D_O^{'2} D_O) + O(A N_f \chi^3 D_O^{'2} D_O) + 
O(A N_f^2 \chi^3 D_O^{'3}) +  O(A \chi^3 D_S^3)+
O(A N_f \chi^3 D_O^{'2} D_S)]$.

In both cases, the cost is linear in the system 
area $A$ as we originally desired.
However, it is instructive to compare these costs 
to an implementation without a PEPO.
In a naive implementation of the exact term-by-term 
contraction of each $n_i n_j$ operator
in the Coulomb potential,  a single term would 
involve a contraction of cost $O(A \chi^3 D_S^3)$ 
with $\chi \sim D_S^2$,
and there would be $O(A^2)$ such terms, giving 
an $O(A^3)$ cost. Assuming a reasonably large 
value for $D_S$, this cost can be compared to the analogous 
term in the (snake) PEPO contraction cost,
which gives an approximate crossover when $A^2 
\sim N_t D_O^3$, which for $N_t=10$, $D_O=6$, 
corresponds to $A\sim 50$.
In a more sophisticated exact implementation, we 
could rewrite $\sum_{ij} V_{ij} n_i n_j$ as 
$\sum_i n_i \hat{O}_i$, with
$\hat{O}_i = \sum_j V_{ij} n_j$. Each $\hat{O}_i$ 
can be represented as a snake-like MPO with bond 
dimension $D=3$, and the cost of contracting
a single $\hat{O}_i$ expectation value is then 
$O(A \chi^3 D_S^3)$ with $\chi \sim D D_S^2$, 
with $O(A)$ such terms. The crossover
with our (snake) PEPO representation then occurs 
when $A \sim  8 N_t$, which for $N_t=10$ corresponds 
to $A < 100$. Thus in either comparison,
a crossover between our PEPO representation and 
other implementations of the long-range operator 
is achievable already at modest
lattice sizes.

%% the computational scaling would be similar to the above expression, but
%% with $N_t=D=D'=N_f=1$ and each term would be multiplied by $A^2$. 
%% {\color{red} An outline of the derivation of these costs is given in Appendix B.}
%% and is thus
%% asymptotically optimal
%% Compared to a naive
%% implementation of the term-by-term by contraction of each $n_i n_j$ operator in the Coulomb potential
%% (which, for each $n_i n_j$  would carry a similar scaling as the above expressions, but with 
%% we would expect the above scheme to become cheaper for $A/2 > N_t D^{'3}D N_f^2$
%% in a naive
%% scheme would 
%% and with each term multiplied by $A$
%% and with each term multiplied by
%% If we assume that the $N_f^2$ term is the dominant term, then we see that 
%% expressions
%% in the system size $A$ is the crucial feature for 
%% making long-range interactions computationally viable.
%% {\color{red} An outline of their derivation is given in Appendix B.}

\begin{figure}[t]
\includegraphics[width=0.225\textwidth]{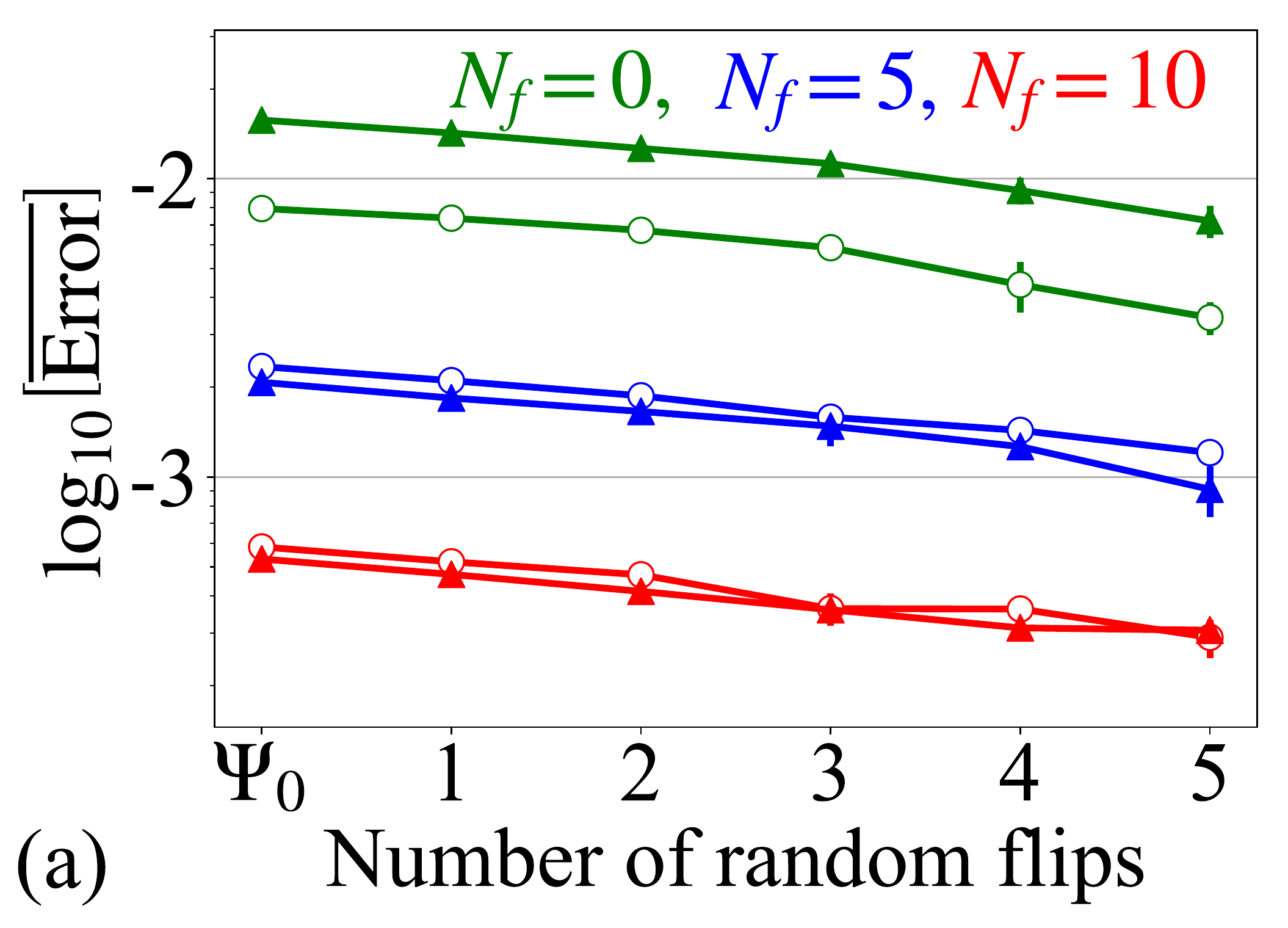} 
\includegraphics[width=0.23\textwidth]{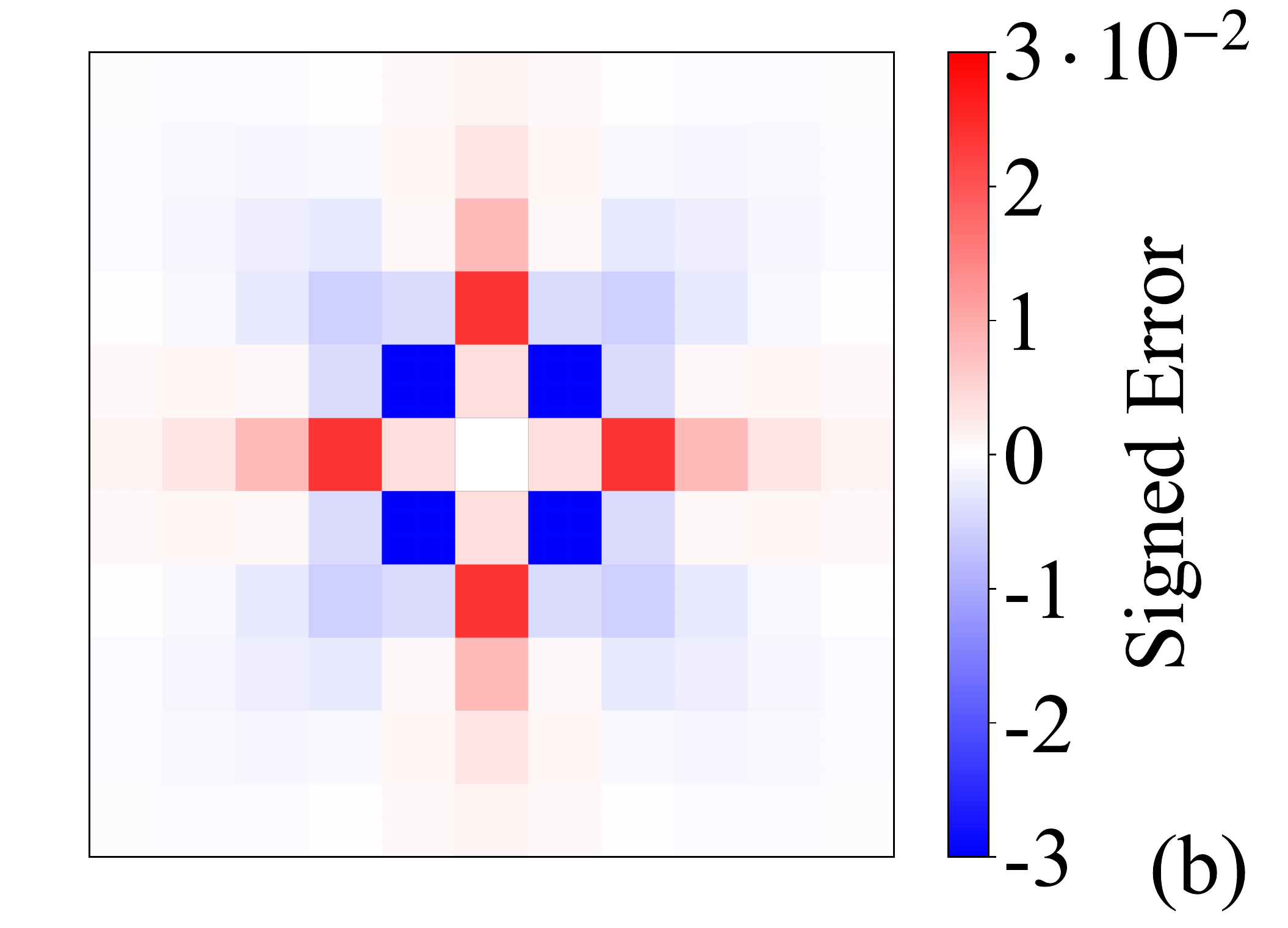}\\[-0.8ex]
\caption{
(a) Average accuracy of energy per site expectation values
for $6\times6$ FM and AFM 
trial PEPS with $D_S=1$. The solid triangular markers show FM states
while the open circles show AFM states. $\Psi_0$ is a true FM or AFM
state, while the ``$x$ flip" regions are $\Psi_0$ perturbed by $x$ 
random spin flips. The average error is taken over 5 PEPS for
each $x$ and each $N_f$. 
(b) The signed error 
$1/r_{ij} - \tilde{V}^{[0]}_{\mathrm{fit}}(r_{ij})$,
where $r_{ij}=0$ is the white square in the center,
each adjacent square is $r_{ij}=1$, etc. 
For (a)-(b) the fitted potentials are obtained from 
Eq. \eqref{eq:scaleFit} with $N_t=12$.}
\label{fig:energy_error}
\end{figure}

\section{V. RESULTS}
To numerically test  our PEPO's faithful discretized representation of
long range interactions, 
we have explicitly constructed a long-range $S$=1/2 Heisenberg
Hamiltonian on $4\times4$, $6\times6$, and $8\times8$ square lattices,
\begin{equation}
\hat{H} = \sum_{i<j} \frac{\vec{S}_i \cdot \vec{S}_j}{r_{ij}},
\label{eqn:Hamiltonian}
\end{equation}
in which every pair of spins has an interaction strength
of Coulomb form. To represent this operator, we first used the 
fitting scheme described in Eq. \eqref{eq:scaleFit} with
$N_t=12$.
Figure \ref{fig:energy_error}(a) shows the accuracy of
the energy per site expectation value ($e_0$)  
for $6\times6$ trial ferromagnetic (FM) and 
anti-ferromagnetic (AFM) PEPS with 
$D_S=1$.
The FM and AFM states show similar levels of error
for a given value of $N_f$, indicating that the
fitted operator can obtain 
similar levels of error even for states which have different structures of the
\textit{signed} error. 
%% . This suggests that the weighted
%% least squares method is a good general technique for fitting
%% the potential $V(r_{ij})$ because it is able to obtain similar
%% levels of error for states which have different structures of the
%% \textit{signed} error. However, the presence of nontrivial signed
%% error cancellation can be seen from the slightly negative slope
%% of all the lines in Figure \ref{fig:energy_error}(a). Fitting
%% techniques which more rigorously account for 
%% signed error cancellation may be
%% able to obtain significantly higher levels of accuracy for a given $N_f$.

\begin{figure}[t]
\begin{center}
\begin{tabular}{c}
\includegraphics[width=0.47\textwidth]{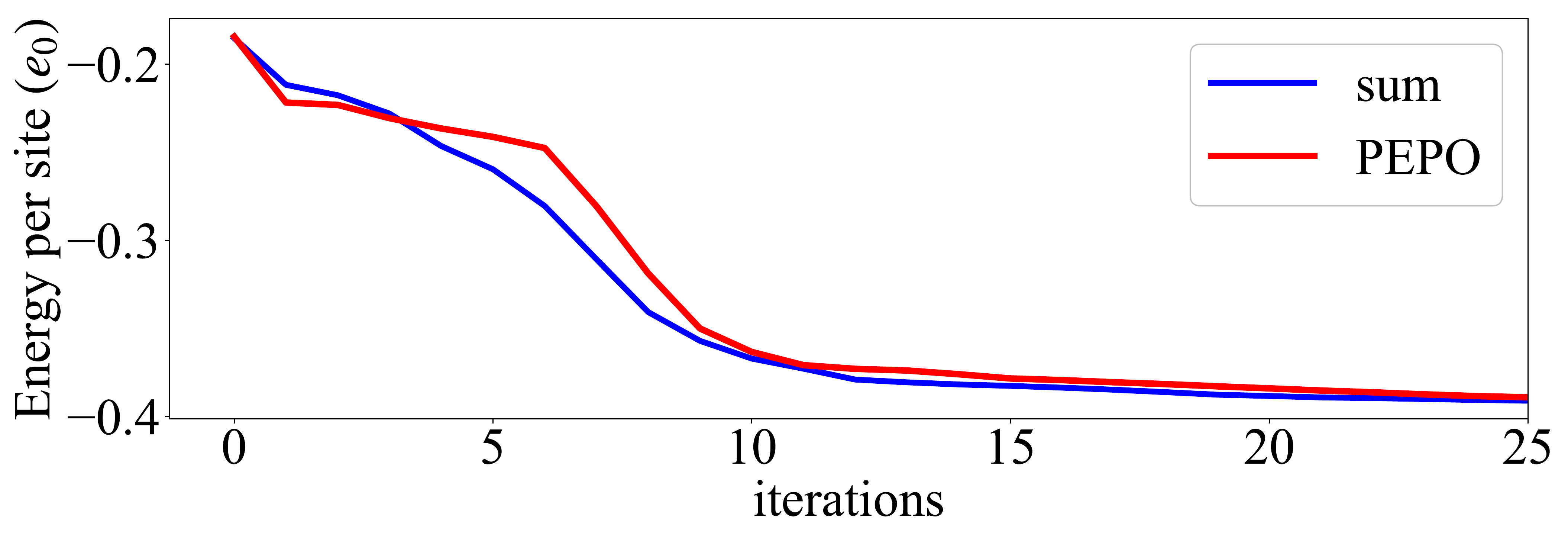} \\
\begin{tabular}{|c|c|c|c|c|}
\hline
 & $\chi$ & sum $e_0$ & PEPO $e_0$ & %direct($\ket{\psi_0^{[P]}}$) &
$ \langle \psi_0^{[P]} | \psi_0^{[s]} \rangle$ \\
\hline
$4\times4$, $D_S=1$ & 40 & -0.184314 &  -0.184425 & %-0.184295 & 
0.999244 \\
$4\times4$, $D_S=2$ & 100 & -0.408209 & -0.408492 & %-0.408160 & 
0.999070 \\
$4\times4$, exact & -- & 0.424577 & -- & -- \\
$8\times8$, $D_S=1$ & 40 & -0.193983 & -0.193861 & %-0.193979 & 
0.994549 \\
$8\times8$, $D_S=2$ & 120 & -0.414653 & -0.414422 & %-0.414489 & 
0.989271\\
$8\times8$, exact & -- & -0.431648 & -- & -- \\
\hline
\end{tabular}
\end{tabular}
\end{center}
\caption{\textit{Top}: The trajectories over the first 25 
iterations of the
energy optimization for the $4\times4$ $D_S=2$ system
using the PEPO and the explicit sum over
all $O(A^2)$ terms in \eqref{eqn:Hamiltonian}. The
long tails of the trajectories are excluded for clarity.
\textit{Bottom}: Ground state energies per site $e_0$ for the 
Hamiltonian \eqref{eqn:Hamiltonian} with various system 
sizes and bond dimensions.
The fifth column is the overlap of the normalized ground states
obtained with the two different methods.
In all cases $N_f=4$ and $N_t=12$.
The ``exact" rows are
the results of converged DMRG calculations.
}
\label{fig:opt}
\end{figure}

We next performed a simple gradient-based 
variational optimization for the
ground state PEPS with $D_S=1, 2$
~\cite{vanderstraeten2016gradient,corboz2016variational}.
Note that our goal here is not to demonstrate fully converged
physics with respect to the PEPS bond dimension, which will be
discussed in future studies, but rather to
show that our PEPO leads to a stable optimization procedure.
Here we refined the fit for each lattice size to ensure that
%% We employed a more accurate but less general 
%% least-squares fitting
%% which obtained separate fitted potentials for each
%% different system size \footnote{
%% Instead of rescaling the fit done over the
%% entire Ising lattice, a unique fit was performed for each 
%% system that accounts for the system size and the number 
%% fictitious sites. By doing this, the fit is performed on
%% only the Ising grid points whose correlation
%% functions are explicitly evaluated for
%% the given system. This reduces the fitting error by
%% avoiding both the additional factor of $N_f$ incurred 
%% in \eqref{eq:scaleFit} and the fitting of extraneous points.
%% However, the penalty for performing the fit in this 
%% way is that it sacrifices the extrapolation properties of
%% Eq. \eqref{eq:scaleFit}.}. 
%% This allowed us to, in all cases, tune
the maximum PEPO
fitting error was limited to $\sim 4.5\cdot10^{-4}$ with 
only $N_f=4$, $N_t=12$.
Fig.~\ref{fig:opt} shows the initial 
convergence behavior of the energy optimization
using the PEPO
compared to the same optimization
using the more expensive sum over terms 
formalism.
We observe that the trajectories are similar
and the use of the PEPO does 
not change the stability of the gradient 
optimization, although it does require a 
larger value of $\chi$. The small-$D_S$
converged energies and normalized wavefunction overlaps
are given in Fig. \ref{fig:opt}. In all cases,
the CF-PEPO nicely reproduces the explicit sum-over-terms algorithm,
as the maximum fitting error
is faithfully reflected in the accuracy of $e_0$. 
It is also interesting to see that the error of the
ground-state energy using $D_S=2$ is $\sim 3\%$ 
for both the $4\times 4$ and $8\times 8$ lattice, suggesting
that the entanglement does not grow significantly with
system size despite the long-range interaction, which is a similar 
observation to other simulations of physical Coulombic systems.
%% Note that these results
%% are not converged with respect to $D_S$, but for a given 
%% $D_S$ the CF-PEPO 

\noindent \textit{Conclusions}. --- In summary, we have
detailed the efficient construction of a PEPO 
capable of encoding long-range
interactions in 2D TNS that maintains the 
strengths of tensor network algorithms: systematically
improvable accuracy and linear computational complexity
in the system size. Despite an increased cost prefactor 
compared to local simulations, this approach allows 
for the possibility of practically including 
long-range interactions
in numerical studies of physically realistic systems that have an
entanglement structure consistent with PEPS. The crossover
  between our approach and other more naive implementations
  of long-range interactions can be achieved at modest system sizes.
  In the context of \textit{ab initio} electronic structure calculations,
  while there remain many issues to explore, in particular associated
  with the continuum limit of relevance to such applications, this advance presents
  a first step towards these calculations using higher dimensional tensor networks.

\begin{acknowledgements}
Primary support for this work was from MURI FA9550-18-1-0095,
which supported MJO. Additional support was from the
US National Science Foundation via grant CHE-1665333
for ZL. GKC acknowledges support from the Simons
Foundation.
\end{acknowledgements}
 
\bibliographystyle{apsrev4-1}
\bibliography{mybib}

\begin{thebibliography}{49}%
\makeatletter
\providecommand \@ifxundefined [1]{%
 \@ifx{#1\undefined}
}%
\providecommand \@ifnum [1]{%
 \ifnum #1\expandafter \@firstoftwo
 \else \expandafter \@secondoftwo
 \fi
}%
\providecommand \@ifx [1]{%
 \ifx #1\expandafter \@firstoftwo
 \else \expandafter \@secondoftwo
 \fi
}%
\providecommand \natexlab [1]{#1}%
\providecommand \enquote  [1]{``#1''}%
\providecommand \bibnamefont  [1]{#1}%
\providecommand \bibfnamefont [1]{#1}%
\providecommand \citenamefont [1]{#1}%
\providecommand \href@noop [0]{\@secondoftwo}%
\providecommand \href [0]{\begingroup \@sanitize@url \@href}%
\providecommand \@href[1]{\@@startlink{#1}\@@href}%
\providecommand \@@href[1]{\endgroup#1\@@endlink}%
\providecommand \@sanitize@url [0]{\catcode `\\12\catcode `\$12\catcode
  `\&12\catcode `\#12\catcode `\^12\catcode `\_12\catcode `\%12\relax}%
\providecommand \@@startlink[1]{}%
\providecommand \@@endlink[0]{}%
\providecommand \url  [0]{\begingroup\@sanitize@url \@url }%
\providecommand \@url [1]{\endgroup\@href {#1}{\urlprefix }}%
\providecommand \urlprefix  [0]{URL }%
\providecommand \Eprint [0]{\href }%
\providecommand \doibase [0]{http://dx.doi.org/}%
\providecommand \selectlanguage [0]{\@gobble}%
\providecommand \bibinfo  [0]{\@secondoftwo}%
\providecommand \bibfield  [0]{\@secondoftwo}%
\providecommand \translation [1]{[#1]}%
\providecommand \BibitemOpen [0]{}%
\providecommand \bibitemStop [0]{}%
\providecommand \bibitemNoStop [0]{.\EOS\space}%
\providecommand \EOS [0]{\spacefactor3000\relax}%
\providecommand \BibitemShut  [1]{\csname bibitem#1\endcsname}%
\let\auto@bib@innerbib\@empty
%</preamble>
\bibitem [{\citenamefont {White}(1992)}]{white1992dmrg}%
  \BibitemOpen
  \bibfield  {author} {\bibinfo {author} {\bibfnamefont {S.~R.}\ \bibnamefont
  {White}},\ }\href@noop {} {\bibfield  {journal} {\bibinfo  {journal}
  {Physical review letters}\ }\textbf {\bibinfo {volume} {69}},\ \bibinfo
  {pages} {2863} (\bibinfo {year} {1992})}\BibitemShut {NoStop}%
\bibitem [{\citenamefont {White}(1993)}]{white1993dmrg}%
  \BibitemOpen
  \bibfield  {author} {\bibinfo {author} {\bibfnamefont {S.~R.}\ \bibnamefont
  {White}},\ }\href@noop {} {\bibfield  {journal} {\bibinfo  {journal}
  {Physical Review B}\ }\textbf {\bibinfo {volume} {48}},\ \bibinfo {pages}
  {10345} (\bibinfo {year} {1993})}\BibitemShut {NoStop}%
\bibitem [{\citenamefont {Vidal}(2007)}]{vidal2007mera}%
  \BibitemOpen
  \bibfield  {author} {\bibinfo {author} {\bibfnamefont {G.}~\bibnamefont
  {Vidal}},\ }\href@noop {} {\bibfield  {journal} {\bibinfo  {journal}
  {Physical review letters}\ }\textbf {\bibinfo {volume} {99}},\ \bibinfo
  {pages} {220405} (\bibinfo {year} {2007})}\BibitemShut {NoStop}%
\bibitem [{\citenamefont {Changlani}\ \emph {et~al.}(2009)\citenamefont
  {Changlani}, \citenamefont {Kinder}, \citenamefont {Umrigar},\ and\
  \citenamefont {Chan}}]{changlani2009approximating}%
  \BibitemOpen
  \bibfield  {author} {\bibinfo {author} {\bibfnamefont {H.~J.}\ \bibnamefont
  {Changlani}}, \bibinfo {author} {\bibfnamefont {J.~M.}\ \bibnamefont
  {Kinder}}, \bibinfo {author} {\bibfnamefont {C.~J.}\ \bibnamefont {Umrigar}},
  \ and\ \bibinfo {author} {\bibfnamefont {G.~K.-L.}\ \bibnamefont {Chan}},\
  }\href@noop {} {\bibfield  {journal} {\bibinfo  {journal} {Physical Review
  B}\ }\textbf {\bibinfo {volume} {80}},\ \bibinfo {pages} {245116} (\bibinfo
  {year} {2009})}\BibitemShut {NoStop}%
\bibitem [{\citenamefont {Mezzacapo}\ \emph {et~al.}(2009)\citenamefont
  {Mezzacapo}, \citenamefont {Schuch}, \citenamefont {Boninsegni},\ and\
  \citenamefont {Cirac}}]{mezzacapo2009ground}%
  \BibitemOpen
  \bibfield  {author} {\bibinfo {author} {\bibfnamefont {F.}~\bibnamefont
  {Mezzacapo}}, \bibinfo {author} {\bibfnamefont {N.}~\bibnamefont {Schuch}},
  \bibinfo {author} {\bibfnamefont {M.}~\bibnamefont {Boninsegni}}, \ and\
  \bibinfo {author} {\bibfnamefont {J.~I.}\ \bibnamefont {Cirac}},\ }\href@noop
  {} {\bibfield  {journal} {\bibinfo  {journal} {New Journal of Physics}\
  }\textbf {\bibinfo {volume} {11}},\ \bibinfo {pages} {083026} (\bibinfo
  {year} {2009})}\BibitemShut {NoStop}%
\bibitem [{\citenamefont {Carleo}\ and\ \citenamefont
  {Troyer}(2017)}]{carleo2017solving}%
  \BibitemOpen
  \bibfield  {author} {\bibinfo {author} {\bibfnamefont {G.}~\bibnamefont
  {Carleo}}\ and\ \bibinfo {author} {\bibfnamefont {M.}~\bibnamefont
  {Troyer}},\ }\href@noop {} {\bibfield  {journal} {\bibinfo  {journal}
  {Science}\ }\textbf {\bibinfo {volume} {355}},\ \bibinfo {pages} {602}
  (\bibinfo {year} {2017})}\BibitemShut {NoStop}%
\bibitem [{\citenamefont {{\"O}stlund}\ and\ \citenamefont
  {Rommer}(1995)}]{ostlund1995thermodynamic}%
  \BibitemOpen
  \bibfield  {author} {\bibinfo {author} {\bibfnamefont {S.}~\bibnamefont
  {{\"O}stlund}}\ and\ \bibinfo {author} {\bibfnamefont {S.}~\bibnamefont
  {Rommer}},\ }\href@noop {} {\bibfield  {journal} {\bibinfo  {journal}
  {Physical review letters}\ }\textbf {\bibinfo {volume} {75}},\ \bibinfo
  {pages} {3537} (\bibinfo {year} {1995})}\BibitemShut {NoStop}%
\bibitem [{\citenamefont {Fannes}\ \emph {et~al.}(1992)\citenamefont {Fannes},
  \citenamefont {Nachtergaele},\ and\ \citenamefont {Werner}}]{fannes1992}%
  \BibitemOpen
  \bibfield  {author} {\bibinfo {author} {\bibfnamefont {M.}~\bibnamefont
  {Fannes}}, \bibinfo {author} {\bibfnamefont {B.}~\bibnamefont
  {Nachtergaele}}, \ and\ \bibinfo {author} {\bibfnamefont {R.~F.}\
  \bibnamefont {Werner}},\ }\href@noop {} {\bibfield  {journal} {\bibinfo
  {journal} {Communications in mathematical physics}\ }\textbf {\bibinfo
  {volume} {144}},\ \bibinfo {pages} {443} (\bibinfo {year}
  {1992})}\BibitemShut {NoStop}%
\bibitem [{\citenamefont {Fannes}\ \emph {et~al.}(1994)\citenamefont {Fannes},
  \citenamefont {Nachtergaele},\ and\ \citenamefont {Werner}}]{fannes1994}%
  \BibitemOpen
  \bibfield  {author} {\bibinfo {author} {\bibfnamefont {M.}~\bibnamefont
  {Fannes}}, \bibinfo {author} {\bibfnamefont {B.}~\bibnamefont
  {Nachtergaele}}, \ and\ \bibinfo {author} {\bibfnamefont {R.}~\bibnamefont
  {Werner}},\ }\href@noop {} {\bibfield  {journal} {\bibinfo  {journal}
  {Journal of functional analysis}\ }\textbf {\bibinfo {volume} {120}},\
  \bibinfo {pages} {511} (\bibinfo {year} {1994})}\BibitemShut {NoStop}%
\bibitem [{\citenamefont {Schollw{\"o}ck}(2011)}]{schollwock2011density}%
  \BibitemOpen
  \bibfield  {author} {\bibinfo {author} {\bibfnamefont {U.}~\bibnamefont
  {Schollw{\"o}ck}},\ }\href@noop {} {\bibfield  {journal} {\bibinfo  {journal}
  {Annals of Physics}\ }\textbf {\bibinfo {volume} {326}},\ \bibinfo {pages}
  {96} (\bibinfo {year} {2011})}\BibitemShut {NoStop}%
\bibitem [{\citenamefont {Nishino}\ and\ \citenamefont
  {Okunishi}(1996)}]{nishino1996corner}%
  \BibitemOpen
  \bibfield  {author} {\bibinfo {author} {\bibfnamefont {T.}~\bibnamefont
  {Nishino}}\ and\ \bibinfo {author} {\bibfnamefont {K.}~\bibnamefont
  {Okunishi}},\ }\href@noop {} {\bibfield  {journal} {\bibinfo  {journal}
  {Journal of the Physical Society of Japan}\ }\textbf {\bibinfo {volume}
  {65}},\ \bibinfo {pages} {891} (\bibinfo {year} {1996})}\BibitemShut
  {NoStop}%
\bibitem [{\citenamefont {Verstraete}\ and\ \citenamefont
  {Cirac}(2004)}]{verstraete2004renormalization}%
  \BibitemOpen
  \bibfield  {author} {\bibinfo {author} {\bibfnamefont {F.}~\bibnamefont
  {Verstraete}}\ and\ \bibinfo {author} {\bibfnamefont {J.~I.}\ \bibnamefont
  {Cirac}},\ }\href@noop {} {\bibfield  {journal} {\bibinfo  {journal} {arXiv
  preprint cond-mat/0407066}\ } (\bibinfo {year} {2004})}\BibitemShut {NoStop}%
\bibitem [{\citenamefont {Verstraete}\ \emph {et~al.}(2006)\citenamefont
  {Verstraete}, \citenamefont {Wolf}, \citenamefont {Perez-Garcia},\ and\
  \citenamefont {Cirac}}]{verstraete2006criticality}%
  \BibitemOpen
  \bibfield  {author} {\bibinfo {author} {\bibfnamefont {F.}~\bibnamefont
  {Verstraete}}, \bibinfo {author} {\bibfnamefont {M.~M.}\ \bibnamefont
  {Wolf}}, \bibinfo {author} {\bibfnamefont {D.}~\bibnamefont {Perez-Garcia}},
  \ and\ \bibinfo {author} {\bibfnamefont {J.~I.}\ \bibnamefont {Cirac}},\
  }\href@noop {} {\bibfield  {journal} {\bibinfo  {journal} {Physical review
  letters}\ }\textbf {\bibinfo {volume} {96}},\ \bibinfo {pages} {220601}
  (\bibinfo {year} {2006})}\BibitemShut {NoStop}%
\bibitem [{\citenamefont {Or{\'u}s}(2014)}]{orus2014practical}%
  \BibitemOpen
  \bibfield  {author} {\bibinfo {author} {\bibfnamefont {R.}~\bibnamefont
  {Or{\'u}s}},\ }\href@noop {} {\bibfield  {journal} {\bibinfo  {journal}
  {Annals of Physics}\ }\textbf {\bibinfo {volume} {349}},\ \bibinfo {pages}
  {117} (\bibinfo {year} {2014})}\BibitemShut {NoStop}%
\bibitem [{\citenamefont {Corboz}(2016{\natexlab{a}})}]{corboz2016improved}%
  \BibitemOpen
  \bibfield  {author} {\bibinfo {author} {\bibfnamefont {P.}~\bibnamefont
  {Corboz}},\ }\href@noop {} {\bibfield  {journal} {\bibinfo  {journal}
  {Physical Review B}\ }\textbf {\bibinfo {volume} {93}},\ \bibinfo {pages}
  {045116} (\bibinfo {year} {2016}{\natexlab{a}})}\BibitemShut {NoStop}%
\bibitem [{\citenamefont {Murg}\ \emph {et~al.}(2007)\citenamefont {Murg},
  \citenamefont {Verstraete},\ and\ \citenamefont
  {Cirac}}]{murg2007variational}%
  \BibitemOpen
  \bibfield  {author} {\bibinfo {author} {\bibfnamefont {V.}~\bibnamefont
  {Murg}}, \bibinfo {author} {\bibfnamefont {F.}~\bibnamefont {Verstraete}}, \
  and\ \bibinfo {author} {\bibfnamefont {J.~I.}\ \bibnamefont {Cirac}},\
  }\href@noop {} {\bibfield  {journal} {\bibinfo  {journal} {Physical Review
  A}\ }\textbf {\bibinfo {volume} {75}},\ \bibinfo {pages} {033605} (\bibinfo
  {year} {2007})}\BibitemShut {NoStop}%
\bibitem [{\citenamefont {Or{\'u}s}\ and\ \citenamefont
  {Vidal}(2009)}]{orus2009simulation}%
  \BibitemOpen
  \bibfield  {author} {\bibinfo {author} {\bibfnamefont {R.}~\bibnamefont
  {Or{\'u}s}}\ and\ \bibinfo {author} {\bibfnamefont {G.}~\bibnamefont
  {Vidal}},\ }\href@noop {} {\bibfield  {journal} {\bibinfo  {journal}
  {Physical Review B}\ }\textbf {\bibinfo {volume} {80}},\ \bibinfo {pages}
  {094403} (\bibinfo {year} {2009})}\BibitemShut {NoStop}%
\bibitem [{\citenamefont {Corboz}\ \emph {et~al.}(2014)\citenamefont {Corboz},
  \citenamefont {Rice},\ and\ \citenamefont {Troyer}}]{PhysRevLett.113.046402}%
  \BibitemOpen
  \bibfield  {author} {\bibinfo {author} {\bibfnamefont {P.}~\bibnamefont
  {Corboz}}, \bibinfo {author} {\bibfnamefont {T.~M.}\ \bibnamefont {Rice}}, \
  and\ \bibinfo {author} {\bibfnamefont {M.}~\bibnamefont {Troyer}},\ }\href
  {\doibase 10.1103/PhysRevLett.113.046402} {\bibfield  {journal} {\bibinfo
  {journal} {Phys. Rev. Lett.}\ }\textbf {\bibinfo {volume} {113}},\ \bibinfo
  {pages} {046402} (\bibinfo {year} {2014})}\BibitemShut {NoStop}%
\bibitem [{\citenamefont {Levin}\ and\ \citenamefont
  {Nave}(2007)}]{levin2007tensor}%
  \BibitemOpen
  \bibfield  {author} {\bibinfo {author} {\bibfnamefont {M.}~\bibnamefont
  {Levin}}\ and\ \bibinfo {author} {\bibfnamefont {C.~P.}\ \bibnamefont
  {Nave}},\ }\href@noop {} {\bibfield  {journal} {\bibinfo  {journal} {Physical
  review letters}\ }\textbf {\bibinfo {volume} {99}},\ \bibinfo {pages}
  {120601} (\bibinfo {year} {2007})}\BibitemShut {NoStop}%
\bibitem [{\citenamefont {Evenbly}\ and\ \citenamefont
  {Vidal}(2015)}]{evenbly2015tensor}%
  \BibitemOpen
  \bibfield  {author} {\bibinfo {author} {\bibfnamefont {G.}~\bibnamefont
  {Evenbly}}\ and\ \bibinfo {author} {\bibfnamefont {G.}~\bibnamefont
  {Vidal}},\ }\href@noop {} {\bibfield  {journal} {\bibinfo  {journal}
  {Physical review letters}\ }\textbf {\bibinfo {volume} {115}},\ \bibinfo
  {pages} {180405} (\bibinfo {year} {2015})}\BibitemShut {NoStop}%
\bibitem [{\citenamefont {Martin}(2004)}]{martin2004electronic}%
  \BibitemOpen
  \bibfield  {author} {\bibinfo {author} {\bibfnamefont {R.~M.}\ \bibnamefont
  {Martin}},\ }\href@noop {} {\emph {\bibinfo {title} {Electronic structure:
  basic theory and practical methods}}}\ (\bibinfo  {publisher} {Cambridge
  university press},\ \bibinfo {year} {2004})\BibitemShut {NoStop}%
\bibitem [{\citenamefont {Szabo}\ and\ \citenamefont
  {Ostlund}(1996)}]{szabo1996modern}%
  \BibitemOpen
  \bibfield  {author} {\bibinfo {author} {\bibfnamefont {A.}~\bibnamefont
  {Szabo}}\ and\ \bibinfo {author} {\bibfnamefont {N.~S.}\ \bibnamefont
  {Ostlund}},\ }\href@noop {} {\emph {\bibinfo {title} {Modern Quantum
  Chemistry: Intro to Advanced Electronic Structure Theory}}}\ (\bibinfo
  {publisher} {Dover publications},\ \bibinfo {year} {1996})\BibitemShut
  {NoStop}%
\bibitem [{\citenamefont {White}\ and\ \citenamefont
  {Martin}(1999)}]{white1999ab}%
  \BibitemOpen
  \bibfield  {author} {\bibinfo {author} {\bibfnamefont {S.~R.}\ \bibnamefont
  {White}}\ and\ \bibinfo {author} {\bibfnamefont {R.~L.}\ \bibnamefont
  {Martin}},\ }\href@noop {} {\bibfield  {journal} {\bibinfo  {journal} {The
  Journal of chemical physics}\ }\textbf {\bibinfo {volume} {110}},\ \bibinfo
  {pages} {4127} (\bibinfo {year} {1999})}\BibitemShut {NoStop}%
\bibitem [{\citenamefont {Chan}\ \emph {et~al.}(2016)\citenamefont {Chan},
  \citenamefont {Keselman}, \citenamefont {Nakatani}, \citenamefont {Li},\ and\
  \citenamefont {White}}]{chan2016matrix}%
  \BibitemOpen
  \bibfield  {author} {\bibinfo {author} {\bibfnamefont {G.~K.-L.}\
  \bibnamefont {Chan}}, \bibinfo {author} {\bibfnamefont {A.}~\bibnamefont
  {Keselman}}, \bibinfo {author} {\bibfnamefont {N.}~\bibnamefont {Nakatani}},
  \bibinfo {author} {\bibfnamefont {Z.}~\bibnamefont {Li}}, \ and\ \bibinfo
  {author} {\bibfnamefont {S.~R.}\ \bibnamefont {White}},\ }\href@noop {}
  {\bibfield  {journal} {\bibinfo  {journal} {The Journal of chemical physics}\
  }\textbf {\bibinfo {volume} {145}},\ \bibinfo {pages} {014102} (\bibinfo
  {year} {2016})}\BibitemShut {NoStop}%
\bibitem [{\citenamefont {Stoudenmire}\ \emph {et~al.}(2012)\citenamefont
  {Stoudenmire}, \citenamefont {Wagner}, \citenamefont {White},\ and\
  \citenamefont {Burke}}]{stoudenmire2012one}%
  \BibitemOpen
  \bibfield  {author} {\bibinfo {author} {\bibfnamefont {E.}~\bibnamefont
  {Stoudenmire}}, \bibinfo {author} {\bibfnamefont {L.~O.}\ \bibnamefont
  {Wagner}}, \bibinfo {author} {\bibfnamefont {S.~R.}\ \bibnamefont {White}}, \
  and\ \bibinfo {author} {\bibfnamefont {K.}~\bibnamefont {Burke}},\
  }\href@noop {} {\bibfield  {journal} {\bibinfo  {journal} {Physical review
  letters}\ }\textbf {\bibinfo {volume} {109}},\ \bibinfo {pages} {056402}
  (\bibinfo {year} {2012})}\BibitemShut {NoStop}%
\bibitem [{\citenamefont {Wagner}\ \emph {et~al.}(2012)\citenamefont {Wagner},
  \citenamefont {Stoudenmire}, \citenamefont {Burke},\ and\ \citenamefont
  {White}}]{wagner2012reference}%
  \BibitemOpen
  \bibfield  {author} {\bibinfo {author} {\bibfnamefont {L.~O.}\ \bibnamefont
  {Wagner}}, \bibinfo {author} {\bibfnamefont {E.}~\bibnamefont {Stoudenmire}},
  \bibinfo {author} {\bibfnamefont {K.}~\bibnamefont {Burke}}, \ and\ \bibinfo
  {author} {\bibfnamefont {S.~R.}\ \bibnamefont {White}},\ }\href@noop {}
  {\bibfield  {journal} {\bibinfo  {journal} {Physical Chemistry Chemical
  Physics}\ }\textbf {\bibinfo {volume} {14}},\ \bibinfo {pages} {8581}
  (\bibinfo {year} {2012})}\BibitemShut {NoStop}%
\bibitem [{\citenamefont {Stoudenmire}\ and\ \citenamefont
  {White}(2017)}]{stoudenmire2017sliced}%
  \BibitemOpen
  \bibfield  {author} {\bibinfo {author} {\bibfnamefont {E.~M.}\ \bibnamefont
  {Stoudenmire}}\ and\ \bibinfo {author} {\bibfnamefont {S.~R.}\ \bibnamefont
  {White}},\ }\href@noop {} {\bibfield  {journal} {\bibinfo  {journal}
  {Physical review letters}\ }\textbf {\bibinfo {volume} {119}},\ \bibinfo
  {pages} {046401} (\bibinfo {year} {2017})}\BibitemShut {NoStop}%
\bibitem [{\citenamefont {Dolfi}\ \emph {et~al.}(2012)\citenamefont {Dolfi},
  \citenamefont {Bauer}, \citenamefont {Troyer},\ and\ \citenamefont
  {Ristivojevic}}]{dolfi2012multigrid}%
  \BibitemOpen
  \bibfield  {author} {\bibinfo {author} {\bibfnamefont {M.}~\bibnamefont
  {Dolfi}}, \bibinfo {author} {\bibfnamefont {B.}~\bibnamefont {Bauer}},
  \bibinfo {author} {\bibfnamefont {M.}~\bibnamefont {Troyer}}, \ and\ \bibinfo
  {author} {\bibfnamefont {Z.}~\bibnamefont {Ristivojevic}},\ }\href@noop {}
  {\bibfield  {journal} {\bibinfo  {journal} {Physical review letters}\
  }\textbf {\bibinfo {volume} {109}},\ \bibinfo {pages} {020604} (\bibinfo
  {year} {2012})}\BibitemShut {NoStop}%
\bibitem [{\citenamefont {Mardirossian}\ \emph {et~al.}(2018)\citenamefont
  {Mardirossian}, \citenamefont {McClain},\ and\ \citenamefont
  {Chan}}]{narbe2018}%
  \BibitemOpen
  \bibfield  {author} {\bibinfo {author} {\bibfnamefont {N.}~\bibnamefont
  {Mardirossian}}, \bibinfo {author} {\bibfnamefont {J.~D.}\ \bibnamefont
  {McClain}}, \ and\ \bibinfo {author} {\bibfnamefont {G.~K.-L.}\ \bibnamefont
  {Chan}},\ }\href@noop {} {\bibfield  {journal} {\bibinfo  {journal} {The
  Journal of chemical physics}\ }\textbf {\bibinfo {volume} {148}},\ \bibinfo
  {pages} {044106} (\bibinfo {year} {2018})}\BibitemShut {NoStop}%
\bibitem [{\citenamefont {Xie}\ \emph {et~al.}(2014)\citenamefont {Xie},
  \citenamefont {Chen}, \citenamefont {Yu}, \citenamefont {Kong}, \citenamefont
  {Normand},\ and\ \citenamefont {Xiang}}]{PhysRevX.4.011025}%
  \BibitemOpen
  \bibfield  {author} {\bibinfo {author} {\bibfnamefont {Z.~Y.}\ \bibnamefont
  {Xie}}, \bibinfo {author} {\bibfnamefont {J.}~\bibnamefont {Chen}}, \bibinfo
  {author} {\bibfnamefont {J.~F.}\ \bibnamefont {Yu}}, \bibinfo {author}
  {\bibfnamefont {X.}~\bibnamefont {Kong}}, \bibinfo {author} {\bibfnamefont
  {B.}~\bibnamefont {Normand}}, \ and\ \bibinfo {author} {\bibfnamefont
  {T.}~\bibnamefont {Xiang}},\ }\href {\doibase 10.1103/PhysRevX.4.011025}
  {\bibfield  {journal} {\bibinfo  {journal} {Phys. Rev. X}\ }\textbf {\bibinfo
  {volume} {4}},\ \bibinfo {pages} {011025} (\bibinfo {year}
  {2014})}\BibitemShut {NoStop}%
\bibitem [{\citenamefont {Corboz}\ and\ \citenamefont
  {Mila}(2014)}]{PhysRevLett.112.147203}%
  \BibitemOpen
  \bibfield  {author} {\bibinfo {author} {\bibfnamefont {P.}~\bibnamefont
  {Corboz}}\ and\ \bibinfo {author} {\bibfnamefont {F.}~\bibnamefont {Mila}},\
  }\href {\doibase 10.1103/PhysRevLett.112.147203} {\bibfield  {journal}
  {\bibinfo  {journal} {Phys. Rev. Lett.}\ }\textbf {\bibinfo {volume} {112}},\
  \bibinfo {pages} {147203} (\bibinfo {year} {2014})}\BibitemShut {NoStop}%
\bibitem [{\citenamefont {Picot}\ and\ \citenamefont
  {Poilblanc}(2015)}]{PhysRevB.91.064415}%
  \BibitemOpen
  \bibfield  {author} {\bibinfo {author} {\bibfnamefont {T.}~\bibnamefont
  {Picot}}\ and\ \bibinfo {author} {\bibfnamefont {D.}~\bibnamefont
  {Poilblanc}},\ }\href {\doibase 10.1103/PhysRevB.91.064415} {\bibfield
  {journal} {\bibinfo  {journal} {Phys. Rev. B}\ }\textbf {\bibinfo {volume}
  {91}},\ \bibinfo {pages} {064415} (\bibinfo {year} {2015})}\BibitemShut
  {NoStop}%
\bibitem [{\citenamefont {Picot}\ \emph {et~al.}(2016)\citenamefont {Picot},
  \citenamefont {Ziegler}, \citenamefont {Or\'us},\ and\ \citenamefont
  {Poilblanc}}]{PhysRevB.93.060407}%
  \BibitemOpen
  \bibfield  {author} {\bibinfo {author} {\bibfnamefont {T.}~\bibnamefont
  {Picot}}, \bibinfo {author} {\bibfnamefont {M.}~\bibnamefont {Ziegler}},
  \bibinfo {author} {\bibfnamefont {R.}~\bibnamefont {Or\'us}}, \ and\ \bibinfo
  {author} {\bibfnamefont {D.}~\bibnamefont {Poilblanc}},\ }\href {\doibase
  10.1103/PhysRevB.93.060407} {\bibfield  {journal} {\bibinfo  {journal} {Phys.
  Rev. B}\ }\textbf {\bibinfo {volume} {93}},\ \bibinfo {pages} {060407}
  (\bibinfo {year} {2016})}\BibitemShut {NoStop}%
\bibitem [{\citenamefont {Zheng}\ \emph {et~al.}(2017)\citenamefont {Zheng},
  \citenamefont {Chung}, \citenamefont {Corboz}, \citenamefont {Ehlers},
  \citenamefont {Qin}, \citenamefont {Noack}, \citenamefont {Shi},
  \citenamefont {White}, \citenamefont {Zhang},\ and\ \citenamefont
  {Chan}}]{Zheng1155}%
  \BibitemOpen
  \bibfield  {author} {\bibinfo {author} {\bibfnamefont {B.-X.}\ \bibnamefont
  {Zheng}}, \bibinfo {author} {\bibfnamefont {C.-M.}\ \bibnamefont {Chung}},
  \bibinfo {author} {\bibfnamefont {P.}~\bibnamefont {Corboz}}, \bibinfo
  {author} {\bibfnamefont {G.}~\bibnamefont {Ehlers}}, \bibinfo {author}
  {\bibfnamefont {M.-P.}\ \bibnamefont {Qin}}, \bibinfo {author} {\bibfnamefont
  {R.~M.}\ \bibnamefont {Noack}}, \bibinfo {author} {\bibfnamefont
  {H.}~\bibnamefont {Shi}}, \bibinfo {author} {\bibfnamefont {S.~R.}\
  \bibnamefont {White}}, \bibinfo {author} {\bibfnamefont {S.}~\bibnamefont
  {Zhang}}, \ and\ \bibinfo {author} {\bibfnamefont {G.~K.-L.}\ \bibnamefont
  {Chan}},\ }\href {\doibase 10.1126/science.aam7127} {\bibfield  {journal}
  {\bibinfo  {journal} {Science}\ }\textbf {\bibinfo {volume} {358}},\ \bibinfo
  {pages} {1155} (\bibinfo {year} {2017})}\BibitemShut {NoStop}%
\bibitem [{\citenamefont {Fano}\ \emph {et~al.}(1998)\citenamefont {Fano},
  \citenamefont {Ortolani},\ and\ \citenamefont {Ziosi}}]{fano1998density}%
  \BibitemOpen
  \bibfield  {author} {\bibinfo {author} {\bibfnamefont {G.}~\bibnamefont
  {Fano}}, \bibinfo {author} {\bibfnamefont {F.}~\bibnamefont {Ortolani}}, \
  and\ \bibinfo {author} {\bibfnamefont {L.}~\bibnamefont {Ziosi}},\
  }\href@noop {} {\bibfield  {journal} {\bibinfo  {journal} {The Journal of
  chemical physics}\ }\textbf {\bibinfo {volume} {108}},\ \bibinfo {pages}
  {9246} (\bibinfo {year} {1998})}\BibitemShut {NoStop}%
\bibitem [{\citenamefont {Chan}\ and\ \citenamefont
  {Head-Gordon}(2002)}]{chan2002highly}%
  \BibitemOpen
  \bibfield  {author} {\bibinfo {author} {\bibfnamefont {G.~K.-L.}\
  \bibnamefont {Chan}}\ and\ \bibinfo {author} {\bibfnamefont {M.}~\bibnamefont
  {Head-Gordon}},\ }\href@noop {} {\bibfield  {journal} {\bibinfo  {journal}
  {The Journal of chemical physics}\ }\textbf {\bibinfo {volume} {116}},\
  \bibinfo {pages} {4462} (\bibinfo {year} {2002})}\BibitemShut {NoStop}%
\bibitem [{\citenamefont {Chan}\ and\ \citenamefont
  {Sharma}(2011)}]{chan2011density}%
  \BibitemOpen
  \bibfield  {author} {\bibinfo {author} {\bibfnamefont {G.~K.-L.}\
  \bibnamefont {Chan}}\ and\ \bibinfo {author} {\bibfnamefont {S.}~\bibnamefont
  {Sharma}},\ }\href@noop {} {\bibfield  {journal} {\bibinfo  {journal} {Annual
  review of physical chemistry}\ }\textbf {\bibinfo {volume} {62}},\ \bibinfo
  {pages} {465} (\bibinfo {year} {2011})}\BibitemShut {NoStop}%
\bibitem [{\citenamefont {Pirvu}\ \emph {et~al.}(2010)\citenamefont {Pirvu},
  \citenamefont {Murg}, \citenamefont {Cirac},\ and\ \citenamefont
  {Verstraete}}]{pirvu2010matrix}%
  \BibitemOpen
  \bibfield  {author} {\bibinfo {author} {\bibfnamefont {B.}~\bibnamefont
  {Pirvu}}, \bibinfo {author} {\bibfnamefont {V.}~\bibnamefont {Murg}},
  \bibinfo {author} {\bibfnamefont {J.~I.}\ \bibnamefont {Cirac}}, \ and\
  \bibinfo {author} {\bibfnamefont {F.}~\bibnamefont {Verstraete}},\
  }\href@noop {} {\bibfield  {journal} {\bibinfo  {journal} {New Journal of
  Physics}\ }\textbf {\bibinfo {volume} {12}},\ \bibinfo {pages} {025012}
  (\bibinfo {year} {2010})}\BibitemShut {NoStop}%
\bibitem [{\citenamefont {Fr{\"o}wis}\ \emph {et~al.}(2010)\citenamefont
  {Fr{\"o}wis}, \citenamefont {Nebendahl},\ and\ \citenamefont
  {D{\"u}r}}]{frowis2010tensor}%
  \BibitemOpen
  \bibfield  {author} {\bibinfo {author} {\bibfnamefont {F.}~\bibnamefont
  {Fr{\"o}wis}}, \bibinfo {author} {\bibfnamefont {V.}~\bibnamefont
  {Nebendahl}}, \ and\ \bibinfo {author} {\bibfnamefont {W.}~\bibnamefont
  {D{\"u}r}},\ }\href@noop {} {\bibfield  {journal} {\bibinfo  {journal}
  {Physical Review A}\ }\textbf {\bibinfo {volume} {81}},\ \bibinfo {pages}
  {062337} (\bibinfo {year} {2010})}\BibitemShut {NoStop}%
\bibitem [{\citenamefont {Crosswhite}\ and\ \citenamefont
  {Bacon}(2008)}]{crosswhite2008finite}%
  \BibitemOpen
  \bibfield  {author} {\bibinfo {author} {\bibfnamefont {G.~M.}\ \bibnamefont
  {Crosswhite}}\ and\ \bibinfo {author} {\bibfnamefont {D.}~\bibnamefont
  {Bacon}},\ }\href@noop {} {\bibfield  {journal} {\bibinfo  {journal}
  {Physical Review A}\ }\textbf {\bibinfo {volume} {78}},\ \bibinfo {pages}
  {012356} (\bibinfo {year} {2008})}\BibitemShut {NoStop}%
\bibitem [{\citenamefont {Crosswhite}\ \emph {et~al.}(2008)\citenamefont
  {Crosswhite}, \citenamefont {Doherty},\ and\ \citenamefont
  {Vidal}}]{crosswhite2008applying}%
  \BibitemOpen
  \bibfield  {author} {\bibinfo {author} {\bibfnamefont {G.~M.}\ \bibnamefont
  {Crosswhite}}, \bibinfo {author} {\bibfnamefont {A.~C.}\ \bibnamefont
  {Doherty}}, \ and\ \bibinfo {author} {\bibfnamefont {G.}~\bibnamefont
  {Vidal}},\ }\href@noop {} {\bibfield  {journal} {\bibinfo  {journal}
  {Physical Review B}\ }\textbf {\bibinfo {volume} {78}},\ \bibinfo {pages}
  {035116} (\bibinfo {year} {2008})}\BibitemShut {NoStop}%
\bibitem [{\citenamefont {Zhao}\ \emph {et~al.}(2010)\citenamefont {Zhao},
  \citenamefont {Xie}, \citenamefont {Chen}, \citenamefont {Wei}, \citenamefont
  {Cai},\ and\ \citenamefont {Xiang}}]{zhao2010renormalization}%
  \BibitemOpen
  \bibfield  {author} {\bibinfo {author} {\bibfnamefont {H.}~\bibnamefont
  {Zhao}}, \bibinfo {author} {\bibfnamefont {Z.}~\bibnamefont {Xie}}, \bibinfo
  {author} {\bibfnamefont {Q.}~\bibnamefont {Chen}}, \bibinfo {author}
  {\bibfnamefont {Z.}~\bibnamefont {Wei}}, \bibinfo {author} {\bibfnamefont
  {J.}~\bibnamefont {Cai}}, \ and\ \bibinfo {author} {\bibfnamefont
  {T.}~\bibnamefont {Xiang}},\ }\href@noop {} {\bibfield  {journal} {\bibinfo
  {journal} {Physical Review B}\ }\textbf {\bibinfo {volume} {81}},\ \bibinfo
  {pages} {174411} (\bibinfo {year} {2010})}\BibitemShut {NoStop}%
\bibitem [{\citenamefont {Baxter}(1982)}]{baxter1982exactly}%
  \BibitemOpen
  \bibfield  {author} {\bibinfo {author} {\bibfnamefont {R.~J.}\ \bibnamefont
  {Baxter}},\ }\href@noop {} {\emph {\bibinfo {title} {Exactly solved models in
  statistical mechanics}}}\ (\bibinfo  {publisher} {Academic Press},\ \bibinfo
  {year} {1982})\ Chap.\ \bibinfo {chapter} {2,7}\BibitemShut {NoStop}%
\bibitem [{\citenamefont {Corboz}\ \emph {et~al.}(2010)\citenamefont {Corboz},
  \citenamefont {Or{\'u}s}, \citenamefont {Bauer},\ and\ \citenamefont
  {Vidal}}]{corboz2010simulation}%
  \BibitemOpen
  \bibfield  {author} {\bibinfo {author} {\bibfnamefont {P.}~\bibnamefont
  {Corboz}}, \bibinfo {author} {\bibfnamefont {R.}~\bibnamefont {Or{\'u}s}},
  \bibinfo {author} {\bibfnamefont {B.}~\bibnamefont {Bauer}}, \ and\ \bibinfo
  {author} {\bibfnamefont {G.}~\bibnamefont {Vidal}},\ }\href@noop {}
  {\bibfield  {journal} {\bibinfo  {journal} {Physical Review B}\ }\textbf
  {\bibinfo {volume} {81}},\ \bibinfo {pages} {165104} (\bibinfo {year}
  {2010})}\BibitemShut {NoStop}%
\bibitem [{\citenamefont {Jordan}\ \emph {et~al.}(2008)\citenamefont {Jordan},
  \citenamefont {Or{\'u}s}, \citenamefont {Vidal}, \citenamefont {Verstraete},\
  and\ \citenamefont {Cirac}}]{jordan2008classical}%
  \BibitemOpen
  \bibfield  {author} {\bibinfo {author} {\bibfnamefont {J.}~\bibnamefont
  {Jordan}}, \bibinfo {author} {\bibfnamefont {R.}~\bibnamefont {Or{\'u}s}},
  \bibinfo {author} {\bibfnamefont {G.}~\bibnamefont {Vidal}}, \bibinfo
  {author} {\bibfnamefont {F.}~\bibnamefont {Verstraete}}, \ and\ \bibinfo
  {author} {\bibfnamefont {J.~I.}\ \bibnamefont {Cirac}},\ }\href@noop {}
  {\bibfield  {journal} {\bibinfo  {journal} {Physical review letters}\
  }\textbf {\bibinfo {volume} {101}},\ \bibinfo {pages} {250602} (\bibinfo
  {year} {2008})}\BibitemShut {NoStop}%
\bibitem [{\citenamefont {Braess}\ and\ \citenamefont
  {Hackbusch}(2005)}]{constNumFitFuncs}%
  \BibitemOpen
  \bibfield  {author} {\bibinfo {author} {\bibfnamefont {D.}~\bibnamefont
  {Braess}}\ and\ \bibinfo {author} {\bibfnamefont {W.}~\bibnamefont
  {Hackbusch}},\ }\href@noop {} {\bibfield  {journal} {\bibinfo  {journal} {IMA
  journal of numerical analysis}\ }\textbf {\bibinfo {volume} {25}},\ \bibinfo
  {pages} {685} (\bibinfo {year} {2005})}\BibitemShut {NoStop}%
\bibitem [{\citenamefont {Xie}\ \emph {et~al.}(2017)\citenamefont {Xie},
  \citenamefont {Liao}, \citenamefont {Huang}, \citenamefont {Xie},
  \citenamefont {Chen}, \citenamefont {Liu},\ and\ \citenamefont
  {Xiang}}]{xie2017optimized}%
  \BibitemOpen
  \bibfield  {author} {\bibinfo {author} {\bibfnamefont {Z.}~\bibnamefont
  {Xie}}, \bibinfo {author} {\bibfnamefont {H.}~\bibnamefont {Liao}}, \bibinfo
  {author} {\bibfnamefont {R.}~\bibnamefont {Huang}}, \bibinfo {author}
  {\bibfnamefont {H.}~\bibnamefont {Xie}}, \bibinfo {author} {\bibfnamefont
  {J.}~\bibnamefont {Chen}}, \bibinfo {author} {\bibfnamefont {Z.}~\bibnamefont
  {Liu}}, \ and\ \bibinfo {author} {\bibfnamefont {T.}~\bibnamefont {Xiang}},\
  }\href@noop {} {\bibfield  {journal} {\bibinfo  {journal} {Physical Review
  B}\ }\textbf {\bibinfo {volume} {96}},\ \bibinfo {pages} {045128} (\bibinfo
  {year} {2017})}\BibitemShut {NoStop}%
\bibitem [{\citenamefont {Vanderstraeten}\ \emph {et~al.}(2016)\citenamefont
  {Vanderstraeten}, \citenamefont {Haegeman}, \citenamefont {Corboz},\ and\
  \citenamefont {Verstraete}}]{vanderstraeten2016gradient}%
  \BibitemOpen
  \bibfield  {author} {\bibinfo {author} {\bibfnamefont {L.}~\bibnamefont
  {Vanderstraeten}}, \bibinfo {author} {\bibfnamefont {J.}~\bibnamefont
  {Haegeman}}, \bibinfo {author} {\bibfnamefont {P.}~\bibnamefont {Corboz}}, \
  and\ \bibinfo {author} {\bibfnamefont {F.}~\bibnamefont {Verstraete}},\
  }\href@noop {} {\bibfield  {journal} {\bibinfo  {journal} {Physical Review
  B}\ }\textbf {\bibinfo {volume} {94}},\ \bibinfo {pages} {155123} (\bibinfo
  {year} {2016})}\BibitemShut {NoStop}%
\bibitem [{\citenamefont {Corboz}(2016{\natexlab{b}})}]{corboz2016variational}%
  \BibitemOpen
  \bibfield  {author} {\bibinfo {author} {\bibfnamefont {P.}~\bibnamefont
  {Corboz}},\ }\href@noop {} {\bibfield  {journal} {\bibinfo  {journal}
  {Physical Review B}\ }\textbf {\bibinfo {volume} {94}},\ \bibinfo {pages}
  {035133} (\bibinfo {year} {2016}{\natexlab{b}})}\BibitemShut {NoStop}%
\end{thebibliography}%

\newpage

\appendix

\section{APPENDIX A: FINITE STATE MACHINE RULES}
The finite state machine picture of a PEPO
views each tensor as a node in a graph, and
each virtual bond of dimension $D$ as a directed 
edge in that graph that can pass $D$ different signals
(or has $D$ different possible states). Note that the
following presentation of these ideas heavily follows
in the spirit of Ref. \cite{frowis2010tensor}.

\noindent \textit{Full 2D FSM}. --- By convention
we have chosen our directed edges to
point up and right so that, for a given tensor at 
position $k$,
its $U$ and $R$ indices pass outgoing signals
while its $D$ and $L$ indices receive incoming
signals. For special combinations of
incoming and outgoing signals for a tensor at 
position $k$, the corresponding tensor entry is a non-zero local operator
$O^{[k]}_{n_k n'_k}$ (which may be the identity operator).
%% that tensor's two physical indices encode a local
%% operator 
%which is either
%a physical operator or the identity operator.
These special
%``certain combinations" of
index values are precisely 
the state machine rules that construct the corresponding
desired state machine. When
the four virtual index values do not match any of 
these desired rules, the value of $O^{[k]}_{n_k n'_k}$
is the zero operator $\hat{\mathbf{0}}$, 
meaning such a configuration of the state machine 
(and therefore such a configuration of the local
operators) is disallowed.
The complete list of rules that define the 
full 2D FSM PEPO which
generates all pairwise interactions 
$\sum_{i<j} \hat{A}_i \hat{B}_j$ with bond
dimension $D=4$ is given in Table \ref{tab:rules}.

\begin{figure*}
\begin{tabular}{cccc}
\includegraphics[width=0.2\textwidth]{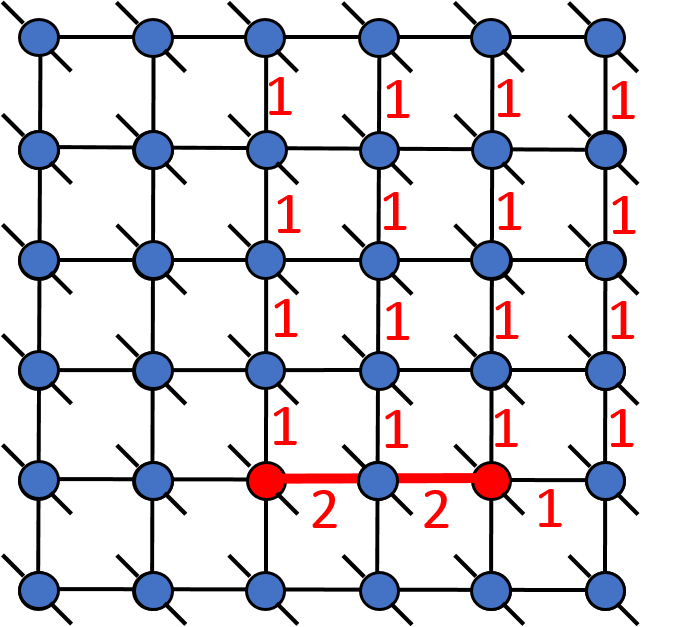} &
\includegraphics[width=0.2\textwidth]{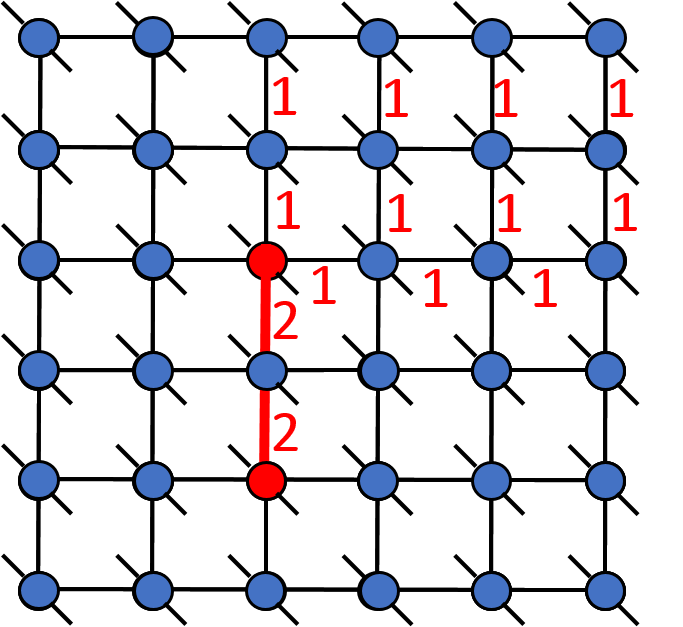} &
\includegraphics[width=0.2\textwidth]{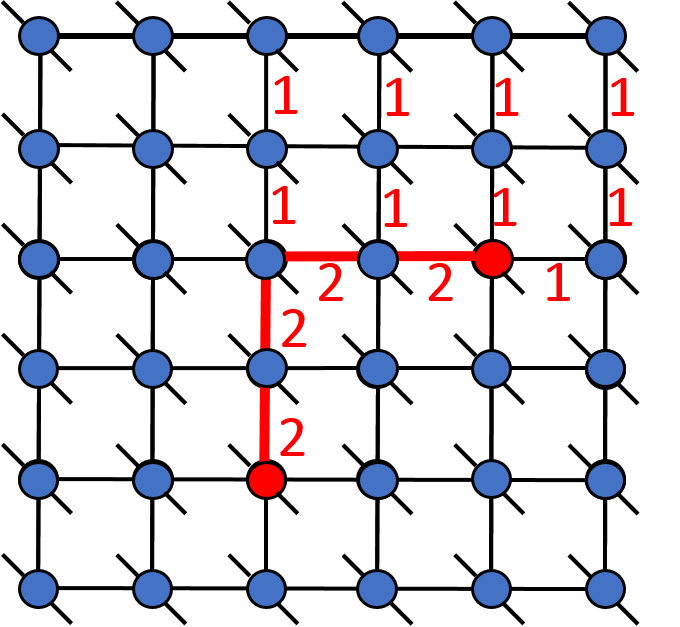} &
\includegraphics[width=0.2\textwidth]{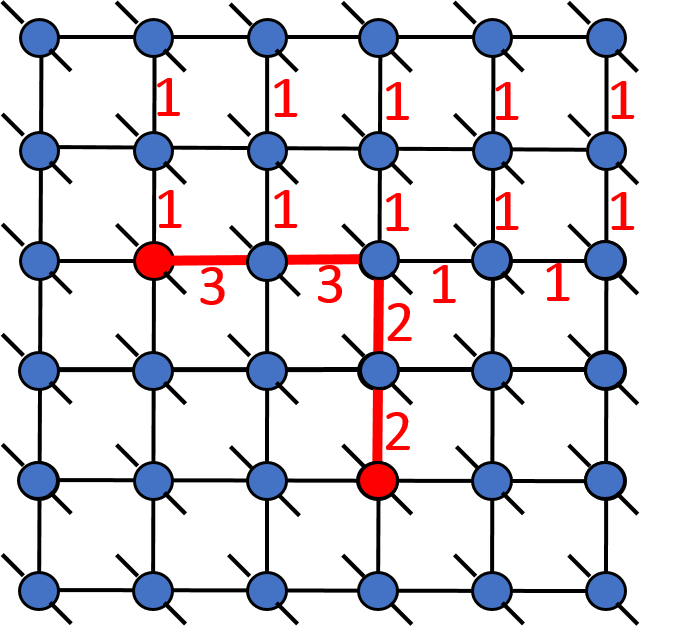} \\
(a) & (b) & (c) & (d)
\end{tabular}
\caption{The four cases 
of rules needed to build the PEPO that encodes 
all the pairwise terms in 
$ \sum_{i < j} \hat{A}_i \hat{B}_j$ for arbitrary
operators $\hat{A}$ and $\hat{B}$. All virtual 
bonds are labeled with their index value, except 
those that are indexed 0 which are left unlabeled. 
The red path denotes the path of the signal from 
$\hat{A}_i$ to $\hat{B}_j$, which are signified 
by the two red tensors. Note that all the blue 
sites will be $\hat{I}$ in these cases.}
\label{fig:rules}
\end{figure*}

Each index value corresponds to a different signal, 
which is used to pass a different message. ``0" is 
the default signal, which generally means that nothing
interesting is happening along that signal path. ``1" is
the signal that tells nearby tensors that they should not
``turn on'' their physical operator $O^{[k]}_{n_k n'_k}$, but
instead should just return the identity operator. This
is used when another tensor along a certain signal path
has turned on its physical operator and does not want an
interaction to be generated along the signal path on which
it just sent a ``1" message. ``2" is the signal that is 
passed along the ``typical" interaction path between the
physical operator at site $i$ and the physical operator at 
site $j$. A typical interaction path is one in which a
signal traveling from site $i$ to site $j$ must
only propagate upward and to the right (along the
allowed directions of the directed edges). The signal
``3" is reserved for the cases in which the signal
traveling from site $i$ to site $j$ must travel to the
left. In order to generate all pairs of sites, one must
either have signals that travel up and to the left or down
and to the right (violating one of the directed edge directions), 
but the case of down and to the left can
be avoided due to the fact that we are generating all
pairs of interactions only once (hence $i<j$ in the 
summations). By convention, we have chosen this pathological
case to be described by a signal that travels up and 
to the left. Since a signal cannot travel against
the direction of a directed edge, this case is resolved
by having the operator at site $j$ (the operator at the ``end"
of the signal) send a ``3" signal to the right, which then
meets with a ``2" signal that was sent upwards from site $i$,
generating an interaction along a ``non-typical" path.
These cases are illustrated diagrammatically
in Fig. \ref{fig:rules}.

\begin{table}[t]
\begin{center}
\begin{tabular}{|c|c|c|}
	\hline
	Rule number & 
\begin{tabular}{c}
Index values \\ $(L_k,U_k,D_k,R_k)$
\end{tabular}
    & $O^{[k]}_{n_k n'_k}$ \\
	\hline
	1 & (0,0,0,0)  & $I_k$ \\
	2 & (0,2,2,0)  & $I_k$ \\
	3 & (2,1,0,2)  & $I_k$ \\
	4 & (0,1,1,0)  & $I_k$ \\
    5 & (1,1,0,1)  & $I_k$ \\
    \hline
    6 & (0,2,0,0)  & $\hat{A}_k$ \\
    7 & (0,1,0,2)  & $\hat{A}_k$ \\
    8 & (0,1,2,2)  & $I_k$ \\
    9 & (0,1,2,1)  & $\hat{B}_k$ \\
    10 & (2,1,0,1)  & $\hat{B}_k$ \\
    \hline
    11 & (3,1,0,3)  & $I_k$ \\
    12 & (3,1,2,1)  & $I_k$ \\
    13 & (0,1,0,3)  & $\hat{B}_k$ \\
    \hline
    $14^*$ & $P^{\text{top right}}_{0,0,0,0}$  & $\hat{0}_k$ \\
	\hline
\end{tabular}
\end{center}
\caption{The rules for the full 2D FSM
PEPO that generates all pairwise 
interactions 
$\sum_{i < j} \hat{A}_i \hat{B}_j$
with $D=4$. All combinations of 
indices not listed in this table correspond to 
$O^{[k]}_{n_k n'_k}=\hat{0}_k$. Importantly, 
$\hat{A}$ and $\hat{B}$
do not have to be the same, although for the 
\textit{ab initio} Hamiltonian under consideration
in the main text, they are both $n_k$. $I_k$ is
simply the identity operator.}
\label{tab:rules}
\end{table}

The rules in Table \ref{tab:rules} are broken up into 
different groups according to what they describe.
Rules 1-5 are background rules that account for
the propagation of ``1" and ``2" signals through
the FSM. Rules 6-10 give the additional rules
necessary for describing a typical interaction.
Rules 11-13 add the rules for non-typical interactions.
Finally, Rule 14 is a special rule that only applies to the
top right tensor in the network, where all signals terminate.
This rule is included to 
disallow the state of the machine where all tensors
have virtual index values $(0,0,0,0)$ and a spurious
1 is added so that the final operator is 
$1+\sum_{i < j} \hat{A}_i \hat{B}_j$ instead of 
$\sum_{i < j} \hat{A}_i \hat{B}_j$.

\noindent \textit{Snake FSM}. --- The snake
construction for the FSM shown in Fig. 1(c) of
the main text is much simpler than the full 2D
FSM above because it is precisely just an MPO
with a few extra dummy legs at each site so that
the direct product with the Ising tensors can
be performed. As discussed briefly in the main text,
the operator-valued local matrices for an MPO
that encodes the interactions $\sum_{i<j} \hat{A}_i
\hat{B}_j$ are given by,
\begin{equation}
M^{[k]} = 
\begin{bmatrix}
\hat{I}_k & \hat{A}_k & \hat{0}_k \\
\hat{0}_k & \hat{I}_k & \hat{B}_k \\
\hat{0}_k & \hat{0}_k & \hat{I}_k
\end{bmatrix}.
\end{equation}

Since this snake imposes an explicit ordering
of all the sites on the 2D square lattice, it
very naturally lends itself to the inclusion
of fermionic statistics at the operator level
via Jordan-Wigner strings. If the operators 
$\hat{A}_i$ and $\hat{B}_j$ are \textit{spinless} 
fermionic creation or annihilation operators
(and $i<j$),
then we have,
\begin{equation}
M^{[k]} = 
\begin{bmatrix}
\hat{I}_k & \hat{a}_k(1-2\hat{n}_k) & \hat{0}_k \\
\hat{0}_k & 1 - 2\hat{n}_k & \hat{b}_k \\
\hat{0}_k & \hat{0}_k & \hat{I}_k
\end{bmatrix},
\label{eq:spinlessJordan}
\end{equation}
where $\hat{a}_k$ and $\hat{b}_k$ are
the hard-core bosonic creation/annihilation operators
and $1-2\hat{n}_k$ encodes the fermionic
statistics. For \textit{spinful} fermionic
operators we have to distinguish between spin
up and spin down
cases. For terms like $\hat{A}_{i\uparrow}
\hat{B}_{j\uparrow}$ we have ,
\begin{equation}
M^{[k]}_{\uparrow\uparrow} = 
\begin{bmatrix}
\hat{I}_k & \hat{a}_k(-1)^{\hat{n}_k} & \hat{0}_k \\
\hat{0}_k & (-1)^{\hat{n}_k} & \hat{b}_k \\
\hat{0}_k & \hat{0}_k & \hat{I}_k
\end{bmatrix},
\label{eq:spinUpJordan}
\end{equation}
and for terms like $\hat{A}_{i\downarrow}
\hat{B}_{j\downarrow}$,
\begin{equation}
M^{[k]}_{\downarrow\downarrow} = 
\begin{bmatrix}
\hat{I}_k & \hat{a}_k & \hat{0}_k \\
\hat{0}_k & (-1)^{\hat{n}_k} & (-1)^{\hat{n}_k}\hat{b}_k \\
\hat{0}_k & \hat{0}_k & \hat{I}_k
\end{bmatrix}.
\label{eq:spinDownJordan}
\end{equation}
Here $1-2\hat{n}_k$ changes to
$(-1)^{\hat{n}_k}$ because we need to
account for the possibility of double 
occupancy at a given site $k$, and this is also
why we distinguish between the spin
up and spin down cases.

\subsection{APPENDIX B: Fitting methodology}

There are many possible ways to fit a given long-range
potential with the correlation functions of an
auxiliary lattice. In this work, we first computed
the Ising model correlation functions at 60 different temperatures.
To choose these temperatures, we first note that away
from the critical temperature of the model ($T_c$),
the correlation functions behave according to $\sim e^{r/\xi}$,
where
\begin{equation}
\xi \propto \left( \frac{T-T_c}{T_c} \right)^{-1},
\end{equation}
is the correlation length. Thus, a geometric series 
in $(T-T_c)/T_c$ was used to select the temperatures,
starting from $T_1 = T_c +\delta$ and ending at 
$T_{60} = 50J/k_B$, where we chose 
$\delta = 5\cdot 10^{-4}$.

%% but in general  is just 
%% a small number that can be tuned. 

\begin{figure}[t]
\begin{center}
\subfloat{
\includegraphics[width=0.47\textwidth]{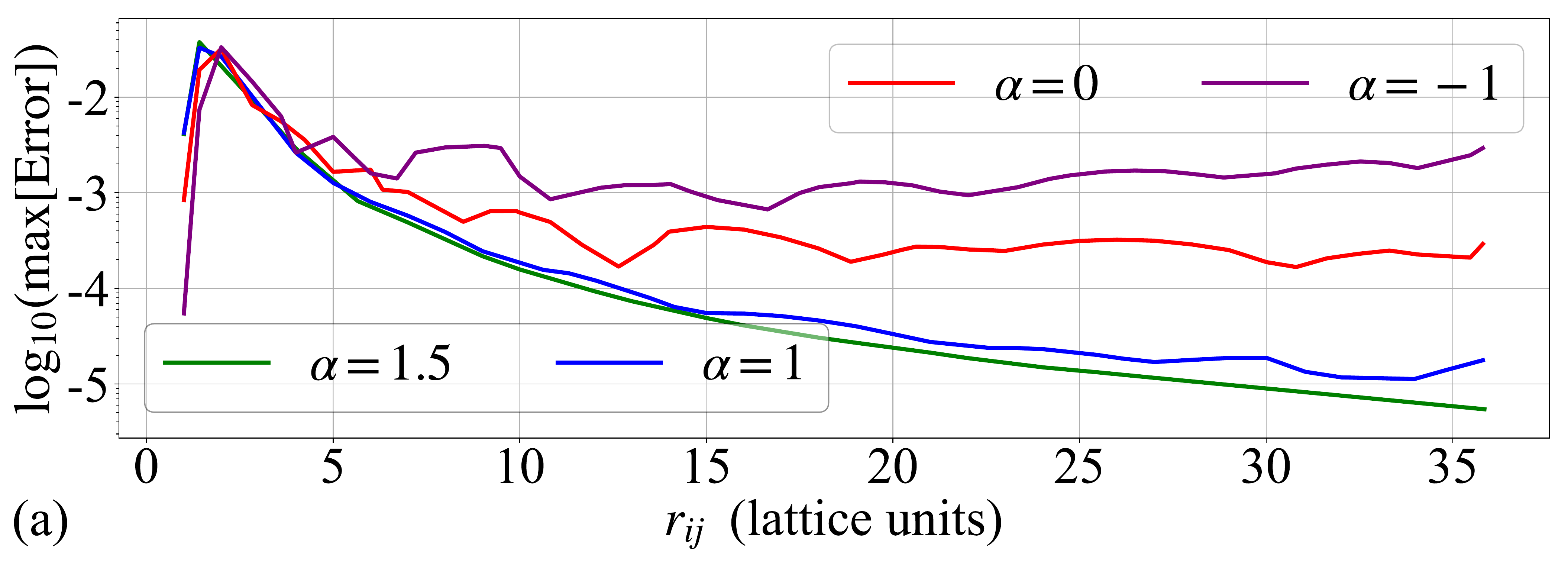}
\label{fig:isotropy1}}\\[-0.8ex]
\subfloat{
\includegraphics[width=0.47\textwidth]{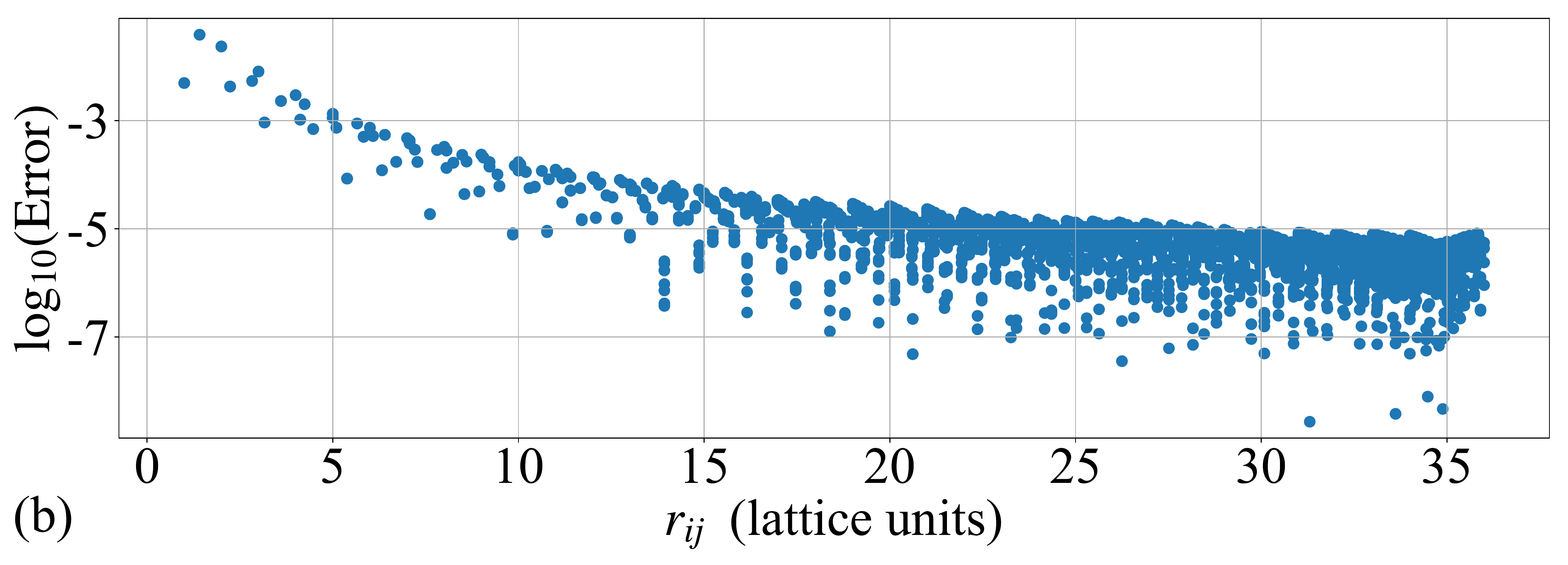}
%\caption{}
\label{fig:isotropy2}}\\[-0.8ex]
\subfloat{
\includegraphics[width=0.47\textwidth]{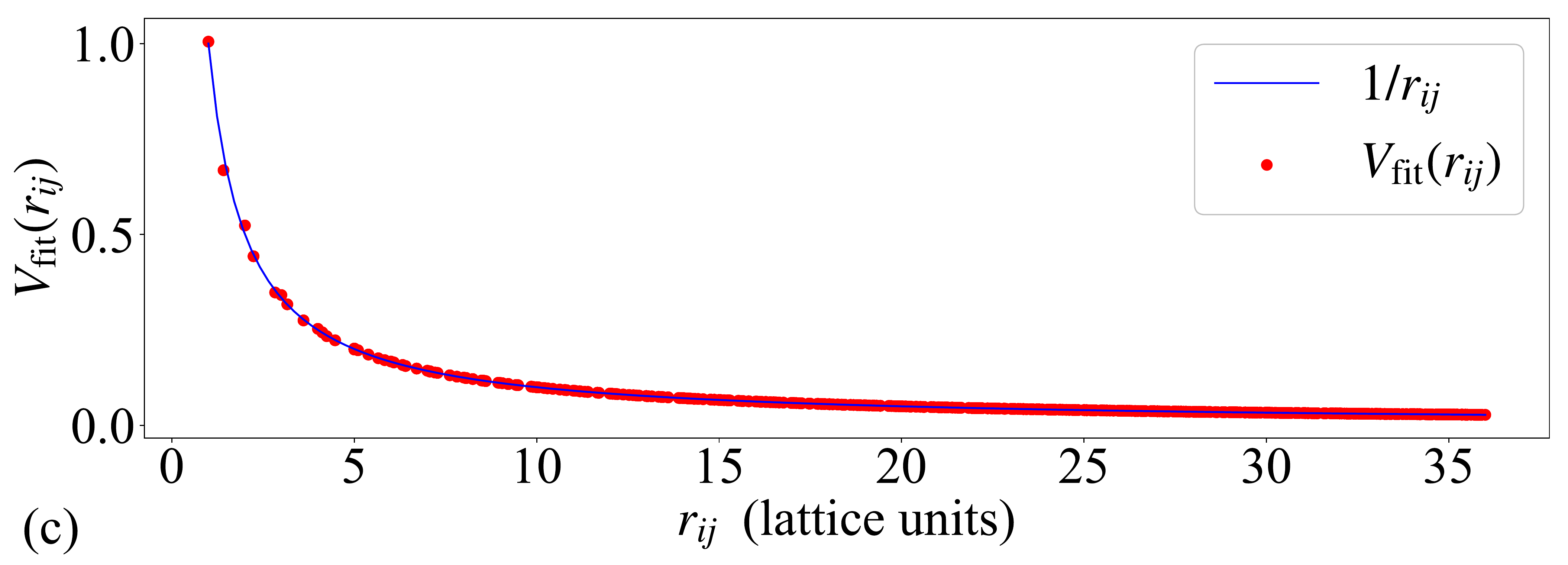}
%\caption{}
\label{fig:isotropy3}}
\\[-2ex]
\caption{(a) The upper envelope of
$\left| V_{\mathrm{fit}}(r_{ij}) - 1/r_{ij} \right|$ 
for different least squares weight functions $r_{ij}^{\alpha}$
with $N_t=12$, $L=199$, and $r_{ij}=R_{ij}$.
(b) All the errors 
$\left| V_{\mathrm{fit}}(r_{ij}) - 1/r_{ij} \right|$
at each $r_{ij}$ for the $N_t=12$, $\alpha=1.5$, 
$L=199$, $r_{ij}=R_{ij}$ fit.
Note that most of the errors for a given $r_{ij}$
are significantly smaller than the upper envelope that
was shown in Fig. \ref{fig:isotropy}a.
(c) The lattice discretized $V_{\mathrm{fit}}(r_{ij})$
compared to the continuous Coulomb potential for the 
$N_t=12$, $\alpha=1.5$, $L=199$, $r_{ij}=R_{ij}$ fit. 
Note that at small
values of $r_{ij}$ the values of 
$V_{\mathrm{fit}}$ visibly deviate from the exact solution,
while as $r_{ij}$ grows the agreement gets
significantly better.
}
\label{fig:fit2}
\end{center}
\end{figure}

With all of this data, a large ``basis matrix'' $\mathbf{A}$
can be formed in which each column is a correlation
function at a different temperature $\beta$.
We then solve the linear regression problem
$\mathbf{A} \vec{c} + \vec{\epsilon} = 1/\vec{r}$, where 
$\vec{c}$ contains the fitting coefficients and
$\vec{\epsilon}$ is the fitting error.
In order to
%reduce the problem size and
improve
conditioning, a rank-revealing QR 
decomposition is performed on $\mathbf{A}$
to give a best guess at the $N_t$
most relevant basis functions (temperatures). 
This allows for a new, smaller matrix 
$\tilde{\mathbf{A}}$ with only $N_t$ columns to
be formed, for which the linear regression
problem is solved by weighted least-squares.
Results of this fitting procedure can be seen in 
Figs. \ref{fig:isotropy} and \ref{fig:fit2}.

%% Weighted least squares fitting appears to be a satisfactory
%% minimization technique for a general Hamiltonian with a
%% long-range potential. However, if more is known
%% \textit{a priori} about the sign structure of the specific
%% interacting operators (eg. $\vec{S}_i$ can take positive and negative
%% values while $n_i$ can only be positive) when they are applied to likely
%% ground states, the signed error can be minimized in order to
%% obtain rigorous error cancellation in the full long-range term of
%% the Hamiltonian.

%% As a final note, in practice we have found the fitting
%% procedure and the accuracy it achieves to be very 
%% sensitive to most parameters aside from $N_t$ and
%% the precise temperatures that are used for the basis
%% functions. For this reason, it is possible that the fitting
%% results presented in the current work do not represent
%% the best solution for all possible system sizes, 
%% choices of $N_f$,
%% and specific Hamiltonians of interest.

\section{APPENDIX C: COMPUTATIONAL COST}
In the main text we claimed that the leading
computational cost for evaluating finite PEPS 
expectation values
using the full 2D FSM CF-PEPO is 
\begin{gather}
N_t [O(A \chi^3 D_O^3)
+ O(A N_f \chi^3 D_O^{'2} D_O)+ \nonumber \\
O(A N_f^2 \chi^3 D_O^{'3}) + O(A \chi^3 D_S^3) + O(A N_f \chi^3 D_O^{'2} D_S)]. \nonumber
\end{gather}
Similarly,
the leading cost of using the snake CF-PEPO was reported
to be, 
\begin{gather}
N_t [O(A \chi^3 D_O^{'2} D_O) + O(A N_f \chi^3 D_O^{'2} D_O) + \nonumber \\
O(A N_f^2 \chi^3 D_O^{'3}) + O(A \chi^3 D_S^3) + O(A N_f \chi^3 D_O^{'2} D_S)], \nonumber
\end{gather}
where in both cases $\chi \sim D_S^2D_O$, $D_O$ is the large PEPO
bond dimension, $D'_O=2$ is the Ising model bond
dimension, and $D_S$ is the PEPS bond dimension.

\begin{figure}[t]
\includegraphics[width=0.47\textwidth]{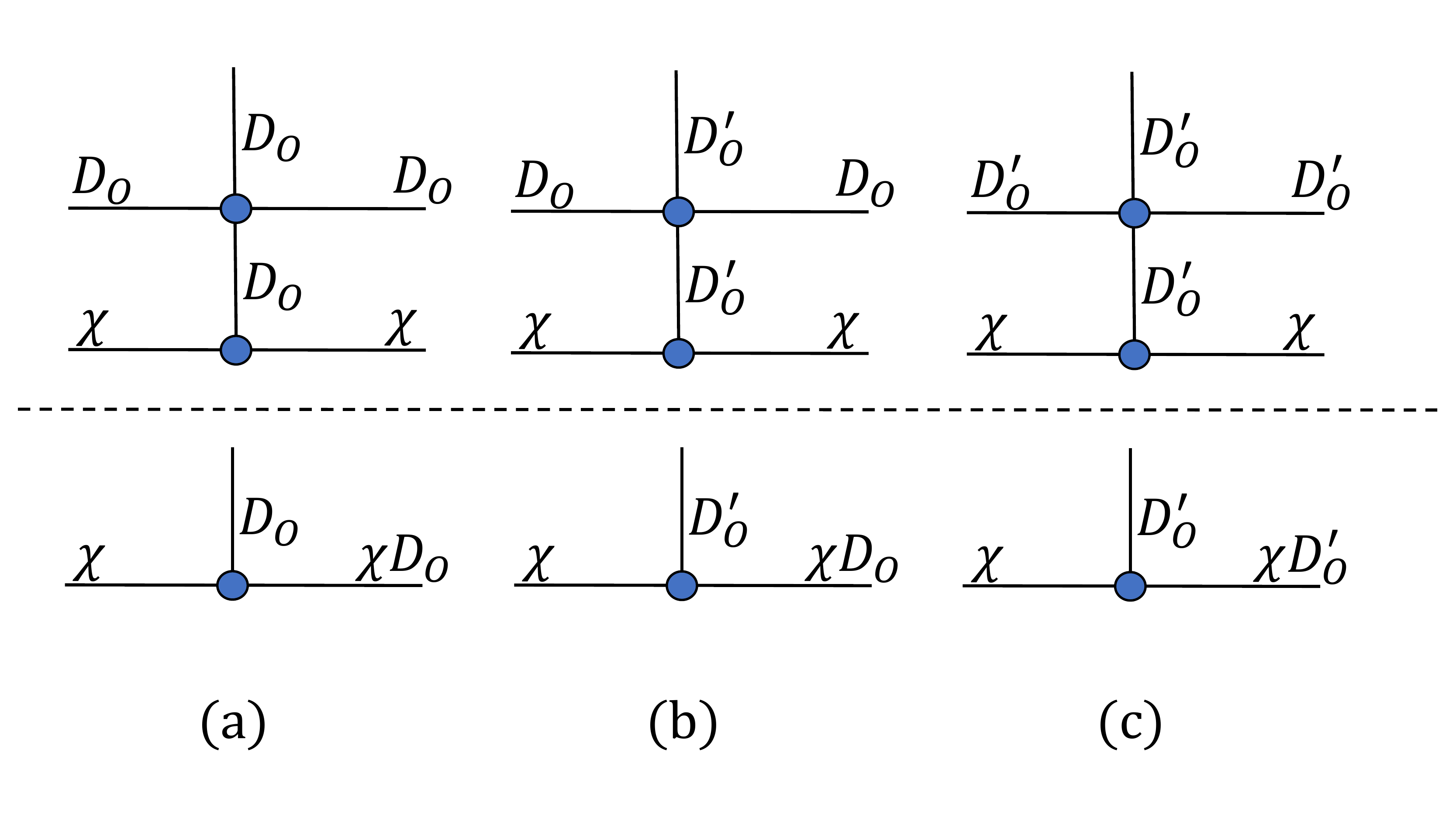}
\caption{Operations which occur during the evaluation of
expectation values using the optimized contraction
scheme. The top row shows contraction of the boundary MPS
into the next row of the grid. The bottom row shows the corresponding 
object on which an SVD must be performed. (a)
operations on physical sites of the PEPS, and also
on physical sites of the full 2D FSM CF-PEPO. (b) operations
on physical sites of the snake CF-PEPO, and also the operations on 
all fictitious or identity tensors which lie in the same row
as the physical PEPO or PEPS tensors. (c) operations on PEPO 
fictitious sites which do not lie in the row or column
of any physical sites.}
\label{fig:compcost}
\end{figure}

In the contraction scheme proposed in \cite{xie2017optimized},
the fundamental operation is to contract a boundary MPS of
bond dimension $\chi$ with a row of tensors corresponding to
either the PEPO layer or the PEPS layer, and then to perform a
subsequent truncation of the boundary bond dimension back to $\chi$. 
%% Ignoring the negligible effect of the physical dimension $d$
%% between the PEPS and PEPO layers, 
The main  
contractions which occur during this process are shown in
the top row of Figure \ref{fig:compcost}. The primary modification
of the scheme in \cite{xie2017optimized} is to account for the
fact that the PEPO has two kinds of sites (fictitious and physical)
which have different bond dimensions. 
For the full 2D FSM
CF-PEPO, (a) shows the contraction of the boundary MPS with 
a physical site tensor in the PEPO; (b) shows the contraction of the boundary MPS
with a fictitious site tensor that falls in the same row as 
physical site PEPO tensors; (c) shows the contraction of the boundary MPS with a fictitious
site tensor that does not fall in the same row or column as the physical
tensors. Diagrams nearly identical to (a) and (c) also occur when contracting
the boundary MPS into the PEPS layer, with the only difference being that $D_O \to D_S$.

The dominant cost arises from the SVDs that must be performed
after contraction to 
%% Subsequently,
%% a DMRG-like sweep of singular value decompositions (SVD) is taken 
%% across the newly formed boundary MPS to
reduce the new composite
bond dimension back to $\chi$. The bottom row of Figure 
\ref{fig:compcost} shows the objects which we need to perform
SVDs on, corresponding to the object that was formed by performing
the contraction right above it in the Figure. The reason why the
objects on the bottom row appear asymmetric along the horizontal
bonds is due to the sweeping nature of the SVDs, which here was assumed
to sweep from left to right.

%% , which has a cost 
%% $O(\chi^2 D^4)$. If we replace $D$ with $D_S$ in the 
%% diagram as well as the cost, then
%% it would show the result of contracting the boundary MPS with
%% a physical site tensor in the PEPS. 
%% (b) shows the contraction of the boundary MPS
%% with a fictitious site tensor that falls in the same row as 
%% physical site PEPO tensors, which has a cost $O(\chi^2 D^2 D^{'2})$.
%% Again, if we replace the $D$ with $D_S$ in both the diagram and
%% the cost, then it would show the results of contracting the 
%% boundary MPS with with a fictious site tensor which lies in
%% the same row as the physical site PEPS tensor. At first glance,
%% this term seems to be extraneous because only the PEPO has 
%% fictitious sites, but in order to make the contraction scheme work,
%% identity tensors with the same dimensions as fictitious site tensors
%% need to be introduced along the bonds of the PEPS
%% as well (but \textit{only} along the bonds of the PEPS, not 
%% creating a completely smaller grid of fictitious sites
%% like in the case of the PEPO).
%% (c) shows the contraction of the boundary MPS with a fictitious
%% site tensor that does not fall in the same row or column as the physical
%% tensors, which has a cost $O(\chi^2 D^{'4})$. As mentioned above,
%% there is no corresponding PEPS term for this part. 

The cost of performing SVDs on these
objects is as follows: (a)$_{\mathrm{PEPO}}$ = $O(\chi^3 D_O^3)$, 
(a)$_{\mathrm{PEPS}}$ = $O(\chi^3 D_S^3)$, 
(b)$_{PEPO}$ = $O(\chi^3  D_O^{'2} D_O)$, (b)$_{PEPS}$ = $O(\chi^3 D_O^{'2} D_S)$,
(c) = $O(\chi^3 D_O^{'3})$, where the subscript denotes whether the boundary 
MPS was first contracted into the PEPS or PEPO layer.

%% Due to the size of $\chi$, it can be seen that the cost of the SVD step
%% always dominates the previous corresponding contraction step.
The operations
of type (a) need to be performed only $O(A)$ times, while the operations of
type (b)
need to be performed $O(A N_f)$ times, and the operations of type (c) 
need to be
performed $O(A N_f^2)$ times. Thus, the total leading cost of evaluating
an expectation value using
the full 2D FSM CF-PEPO is 
\begin{gather}
N_t [O(A \chi^3 D_O^3)
+ O(A N_f \chi^3 D_O^{'2} D_O)+ \nonumber \\
O(A N_f^2 \chi^3 D_O^{'3}) + O(A \chi^3 D_S^3) + O(A N_f \chi^3 D_O^{'2} D_S)]. \nonumber
\end{gather}

To obtain the result for the snake CF-PEPO, one repeats
the above analysis. The only difference is that no operations of
type (a) appear for the PEPO. Instead, the PEPO physical site 
operations have diagrams like type (b). Thus, the first two terms of the 
cost of the snake PEPO look identical, except that one occurs only $O(A)$ times
while the other occurs $O(A N_f)$ times.

\end{document}